\renewcommand{\thefootnote}{\fnsymbol{footnote}}

\documentclass[twocolumn]{aastex631}

\usepackage{enumitem}
\setlist{nosep}
\usepackage{appendix}
\usepackage{bm}
\usepackage{hyperref}

\definecolor{darkgoldenrod}{rgb}{0.72, 0.53, 0.04}

\renewcommand{\thefootnote}{\fnsymbol{footnote}}

\newcommand{\eq}[1]{Eq.~(\ref{#1})}

\newcommand{\Fig}[1]{Fig.~\ref{#1}}

\newcommand{\sect}[1]{Sect.~\ref{#1}}

\newcommand{\msun}{M_{\odot}}
\newcommand{\Msun}{\msun}

\newcommand{\Mbul}{M_{\bullet}}

\newcommand{\Rg}{R_{\rm g}}
\newcommand{\Rhill}{R_{\rm H}}

\newcommand{\Xeff}{\chi_{\rm eff}}
\newcommand{\fret}{f_{\rm ret}}
\newcommand{\tauAGN}{\tau_{\rm AGN}}
\newcommand{\barq}{\overline{q}}
\newcommand{\barXeff}{\overline{\Xeff}}

\newcommand{\emax}{e_{\rm max}}

\newcommand{\sgdefault}{\texttt{sg\_default}}
\newcommand{\tqmdefault}{\texttt{tqm\_default}}

\usepackage{xcolor}

\begin{document}

\title{\texttt{McFACTS} II: Mass Ratio--Effective Spin Relationship of Black Hole Mergers in the AGN Channel}

\author[0000-0001-7163-8712]{Harrison E. Cook}
\affiliation{New Mexico State University, Department of Astronomy, PO Box 30001 MSC 4500, Las Cruces, NM 88003, USA}

\author[0000-0002-9726-0508]{Barry McKernan}
\affiliation{Center for Computational Astrophysics, Flatiron Institute, 
162 5th Ave, New York, NY 10010, USA}
\affiliation{Department of Astrophysics, American Museum of Natural History, New York, NY 10024, USA}
\affiliation{Department of Science, BMCC, City University of New York, New York, NY 10007, USA}
\affiliation{Graduate Center, City University of New York, 365 5th Avenue, New York, NY 10016, USA}

\author[0000-0002-5956-851X]{K.E. Saavik Ford}
\affiliation{Center for Computational Astrophysics, Flatiron Institute, 
162 5th Ave, New York, NY 10010, USA}
\affiliation{Department of Astrophysics, American Museum of Natural History, New York, NY 10024, USA}
\affiliation{Department of Science, BMCC, City University of New York, New York, NY 10007, USA}
\affiliation{Graduate Center, City University of New York, 365 5th Avenue, New York, NY 10016, USA}

\author[0000-0001-7099-765X]{Vera Delfavero}
\affiliation{Gravitational Astrophysics Laboratory, NASA Goddard Space Flight Center, Greenbelt, MD 20771, USA}

\author[0000-0003-2430-9515]{Kaila Nathaniel}
\affiliation{Center for Computational Relativity and Gravitation, Rochester Institute of Technology, Rochester, New York 14623, USA}

\author[0000-0003-0738-8186]{Jake Postiglione}
\affiliation{Department of Astrophysics, American Museum of Natural History, New York, NY 10024, USA}
\affiliation{Graduate Center, City University of New York, 365 5th Avenue, New York, NY 10016, USA}

\author[0009-0005-5038-3171]{Shawn Ray}
\affiliation{Department of Astrophysics, American Museum of Natural History, New York, NY 10024, USA}
\affiliation{Graduate Center, City University of New York, 365 5th Avenue, New York, NY 10016, USA}

\author[0009-0008-5622-6857]{Emily J. McPike}
\affiliation{Department of Astrophysics, American Museum of Natural History, New York, NY 10024, USA}
\affiliation{Graduate Center, City University of New York, 365 5th Avenue, New York, NY 10016, USA}

\author[0000-0001-5832-8517]{Richard O'Shaughnessy}
\affiliation{Center for Computational Relativity and Gravitation, Rochester Institute of Technology, Rochester, New York 14623, USA}

\begin{abstract}

We use the Monte Carlo For AGN (active galactic nucleus) Channel Testing and Simulation (\texttt{McFACTS}\footnote{https://www.github.com/mcfacts/mcfacts}) code to study the effect 
of AGN disk and nuclear star cluster (NSC) parameters on predicted mass distributions for LIGO-Virgo-KAGRA compact binaries forming in AGN disks.
The assumptions we vary include the black hole (BH) initial mass function, disk model, disk size, disk lifetime, NSC model, and the prograde-to-retrograde fraction of newly formed black hole binaries.
Broadly we find that dense, moderately short-lived AGN disks are preferred for producing a $(q,\Xeff)$ anti-correlation like those identified from existing gravitational wave (GW) observations.
Additionally, a BH initial mass function (IMF $\propto M^{-2}$) is preferred over a more top-heavy IMF ($M^{-1}$).
The preferred fraction of prograde-to-retrograde is $>90\%$, to produce results consistent with observations.
\end{abstract}

\section{Introduction} \label{sec:intro}
Binary black hole (BBH) mergers are observed in GW at an increasing rate as LIGO-Virgo-KAGRA (LVK) improve detector sensitivities \citep[e.g.][ see also \footnote{https://gracedb.ligo.org/superevents/public/O4/}]{GW190521astro}. Observed BBH include BH with individual masses in the expected upper mass gap ($\sim 50-120M_{\odot}$) where BH are not expected from stellar evolution \citep{Belcynzski16,Renzo24}. A natural explanation for such BH is hierarchical mergers in a dynamical environment, such as globular clusters \citep{Carl18}, NSC \citep{Antonini&Rasio16,Hoang+18,Fragione+19} and AGN disks \citep{McK14,Bartos17,Stone17, ArcaSedda23,FM25}. Among hierarchical merger models, AGN are notable for both the efficiency of BBH formation as well as for retaining most post-merger products (as even strong kicks from mergers correspond to small orbital perturbations in such a deep potential well). Action of gas as a dynamical coolant and orbital migration producing relatively low velocity dynamical encounters drive BBH formation \citep[e.g.][]{Rowan23,Qian+24,DodiciTremaine24}. The result is a high rate of intermediate mass black hole (IMBH) formation relative to other channels \citep{McK12,Bellovary16,Secunda21,Yang19,McKernan+20a,Hiromichi20,Paola24}, though gap formation may disrupt migration traps that would otherwise facilitate high rates of hierarchical mergers \citep{Gilbaum+25}. The rate of BBH mergers from AGN is also expected to be significantly higher than from quiescent NSCs \citep{Ford22}. 

The observed population of BBH mergers intrinsically favors positive effective spin \citep{Abbot+2023:O3b}, implying formation channels bias spin directions towards alignment with BBH orbital angular momentum. This is not expected from gas-free hierarchical merger channels, but can occur in the AGN channel as BH are torqued towards alignment with the gas disk over time \citep{Hiromichi20spin,Vajpeyi22,McKF24}. Additionally, there is a possible anti-correlation observed in the $(q,\Xeff)$ distribution of BBH mergers \citep{Callister+21}. 
This seems difficult to achieve in standard variants of other dynamical merger channels (e.g. globular clusters, nuclear star clusters), although see \citep{Su25} for a discussion of how such an anti-correlation might arise in a dynamical triple channel. In the isolated binary channel, \citep[e.g.][]{Olejak24,Banerjee24} demonstrate the possibility of generating a ($q,\Xeff$) anti-correlation via stable mass transfer. In the AGN channel, the $(q,\Xeff)$ can arise due to several biases \citep{Mckernan+22:q-Xeff,Santini23, AlexD24}. \citet{Paola24} show that if gas hardening is efficient in AGN, then the $(q,\Xeff)$ distribution should extend to significantly smaller values of $q$ and high $\Xeff$. Thus, LVK O4 results can directly test gas hardening efficiency in AGN channel models.

Note, the AGN channel accounts for a still uncertain fraction ($f_{\rm AGN,BBH}$) of the observed LVK GW events. We will learn a great deal about the average properties of AGN and NSCs out to $z \sim 1$ by constraining $f_{\rm AGN, BBH}$, \emph{regardless of its actual value}, see \citet{FM25} for a more detailed discussion of this point. Thus, in presenting distributions of AGN channel events in ($q,\chi_{\rm eff}$) as a function of model choices, we are not claiming that AGN are responsible for all (or any!) of the observed events. Rather we are demonstrating areas of AGN parameter space that might be ruled out by comparing with LVK O4 (O5) results. This in turn hopefully allows us to constrain $f_{\rm AGN,BBH}$.

Here we present a study using the new \texttt{McFACTS} code \citep[Monte Carlo For AGN Channel Testing and Simulations:][hereafter Paper I]{McKernan+24-McFACTS1} to test the predictions of the AGN channel for the $q$--$\Xeff$ distribution across a wide range of parameter space in both AGN disk and NSC models, with a view to tightly constraining requirements on the AGN channel. \citet[][hereafter Paper III]{Delfavero+24-McFACTS3} presents the first study of the AGN channel in a semi-realistic Universe of AGN and NSCs. \texttt{McFACTS} features substantial improvements to the monte carlo code used in \citet{McKernan+20a} to explore $\Xeff$ of mergers in AGN and \citet{Mckernan+22:q-Xeff} who explored the $q$--$\Xeff$ anti-correlation. The greater flexibility for exploring parameter space includes many improved models for disk-NSC interactions such as dynamical encounters, orbital migration, and binary evolution.

This paper is organized as follows: We provide background information in \sect{sec:background} for the physical processes involved in our model. \sect{sec:methods} describes the methods we use to vary model parameters. We report results in \sect{sec:results}, discuss the possible implications in \sect{sec:discussion}, and conclude in \sect{sec:conclusion}.

\section{Background}
\label{sec:background}
GW observations provide information about each component mass and spin for a given BBH merger.  In this work, we will focus on two specific BBH parameters: the binary mass ratio $q=M_2/M_1\ (M_1>M_2)$, where $M_{1}$ and $M_{2}$ are the primary and secondary BH masses; and effective spin of the BBH as given by 
\begin{equation}
\label{eqn:Xeff}
 \Xeff = \frac{M_{1}\vec{\chi}_{1} + M_{2} \vec{\chi}_{2}}{M_{1}+M_{2}}\cdot \hat{L}_{\rm bin}
\end{equation}
where $\vec{\chi}_{1,2}$ are the dimensionless spin parameters associated with $M_{1},M_{2}$ respectively and $\hat{L}_{\rm bin}$ is the direction of the orbital angular momentum vector of the BBH. 

For intuition in what follows, the range of $q$ among first generation (1g; BH produced via stellar evolution) mergers will depend on the range of the initial mass function (e.g. [10,40] $M_{\odot}$) as well as the power-law of the mass distribution ($\propto M^{-\gamma}$). A steeper mass function with $\gamma = 2$ biases the distribution towards lower mass BH and therefore implies a bias towards $q \sim 1$ as binaries are more likely to form between similar mass BH. A shallower mass function with $\gamma = 1$ permits a higher relative fraction of high mass BH and by extension pairings between a broader range of masses, driving $q$ lower.

Drawing spins ($\vec{\chi}_{1,2}$) from a uniform distribution between those that are aligned and anti-aligned with $\hat{L}_{\rm bin}$ means that the resulting $\Xeff$ distribution should be centered on $\Xeff \sim 0$, with the width of that distribution a function of the initial spin magnitudes $|\vec{\chi}_{1,2}|$; see, e.g., \cite{2021arXiv210409508C} for details. However, a bias towards prograde BBH formation in the AGN channel would favor positive $\Xeff$ values in general. Such a bias might be due to orbital inclination flipping \citep{AlexD24} of an inclined BBH, preferential ionization of retrograde BBH as they spend more time at wider separation \citep[on average,][]{Wang+22,Calcino23}, or some other mechanism. Additionally, once BBH merge, most of their orbital angular momentum as they coalesce will go into natal spin of the newly merged BH, typically $\chi \sim 0.7$ \citep{Tichy08}. Hierarchical mergers containing higher generation (2g or higher) BH that have inherited these large positive spins will therefore strongly favor positive $\Xeff$.

\section{Methods}
\label{sec:methods}
Here we introduce the assumptions, models, and parameters that we vary in the \texttt{McFACTS} code relevant for testing the $q$--$\Xeff$ distribution. We run each of the following setups 100 times to build a statistically significant sample. In each run, our code evaluates the following physical processes: 1) BH populate the disk, 2) gas damping circularizes eccentric BH orbiters until their orbits are sufficiently circular at which point they migrate radially, 3) encounters between circularized BH form BBH whose separations shrink (harden) or expand (soften) due to gas and dynamical processes, 4) sufficiently hardened BH merge, 5) merger recoils put BH remnants on eccentric orbits where the process continues or removes them from the disk, and 6) gas torque acting on disk-crossing BH orbits replenishes the embedded BH population from the NSC. For details of the code and further assumptions made therein see the discussion in Paper I.

\subsection{Disk Models}
As we cannot yet observationally resolve them, the physical conditions of AGN disks are poorly constrained. Instead we rely on models fit to spectroscopic and multi-wavelength variability data. We use two common one-dimensional radial models \citet[][hereafter SG]{SirkoGoodman03} and \citet[][hereafter TQM]{Thompson+05}. The inner regions of the SG disk are likely to be more accurate than the outer regions where assumptions are made to prevent the disk from collapsing due to gravitational instability. In contrast, the star formation models supporting the outer regions of the TQM model are more accurately implemented, but the model's opacity for inner disk annuli are likely too low. For both models we truncate the disk at some outer value $R_{\rm out}$, assume the disk lives in a steady state for time $\tauAGN$, and use the \texttt{pAGN} code to construct our disk models \citep{Gangardt+24-pAGN}.

Analogous to proto-planetary disks, migration traps and anti-traps may form between regions of outward and inward migration \citep{Lyra+10}. We calculate migration torques for circular BH orbiters (singles and binaries) and change their semi-major axes depending on their location relative to user-defined traps according to \citet{Paardekooper+10} as implemented in \citet{GrishinGilbaumStone24} and consider the additional influence of radiative feedback from accretion \citep{HanklaJiangArmitage20} that reduces the strength of inward migration torques and can even turn them outward. Following \citet{Bellovary16}, we set migration traps at 700 $\Rg$ in SG disks and 500 $\Rg$ in TQM disks, practice, mergers also aggregate near $R\sim10^3,\Rg$ (for $M_\bullet = 10^8\,\Msun$) when using migration prescriptions described by \citet{GrishinGilbaumStone24} as BH enter the region of lower migration speeds.

The choice of migration torque model may be important for disk structures around SMBH of different mass, and \texttt{McFACTS} provides the user with two options: a two-dimensional model from \citet{Paardekooper+10} and a three-dimensional (3D) model from \citet{JimenezMassset17}.
In general, the Jim\'enez-Masset 3D torques are weaker across the bulk disk by up to $\sim$30\%.
This tends to drop the overall merger rate in disks (including at the trap/swamp if present).
Likewise migration traps move radially outward in units of $\Rg$ as the SMBH mass drops well below $10^8\,\Msun$, but in Paper I we compared these models and found little change for $M_\bullet = 10^8\,\Msun$ \citep[see Fig. 6 \& 7 of][]{McKernan+24-McFACTS1}, which is also the SMBH mass we find in Paper III that produces the most mergers \citep{Delfavero+24-McFACTS3}.
Thus, we choose the simpler \citet{Paardekooper+10} torque model here and confine our study to disks around SMBH with mass $M_\bullet = 10^8\,\Msun$.

\subsection{Initial Black Hole Properties}
We draw the initial BH masses from a Pareto distribution with $P(M) \propto M^{-\gamma}$ for $10\leq M \leq40\,\Msun$, with an additional Gaussian component centered on $35M_{\odot}$ (see Paper I). We use $\gamma=2$ as the default throughout this paper and vary it to $\gamma=1$. A steeper distribution (larger $\gamma$) results in fewer high mass BH initially, which leads to more mergers between BH of equal mass favoring $q\sim1$ for first generation encounters, whereas a shallower distribution permits binaries between a broader range of masses, driving $q$ lower.

We draw the initial BH spin directions $\hat{\chi}$ from a uniform distribution over the interval $[0,\pi]$ and the spin magnitudes from a Gaussian distribution centered at $\mu_\chi=0$ with standard deviation $\sigma_{\chi} = 0.1$ as the default and test variations to $\sigma_{\chi}=0.02$ and $0.2$. See Paper I for details on these models.

\subsection{Disk Lifetimes}
AGN lifetimes ($\tauAGN$) are thought to lie in the range $\sim 0.1$ Myr to tens of Myr. The lifetime is important for regulating how long each disk process operates on the nuclear system. We choose a default lifetime for SG disk models of $\tauAGN=0.5$ Myr and study variations of $\tauAGN = 0.25$ and 0.75 Myr. Since the TQM disk is generally less dense which reduces the effectiveness of gas interaction processes, we increase the default lifetime to $\tauAGN = 3$ Myr with variations of $\tauAGN = 2.5$ and 5 Myr.

\subsection{Disk Size}
The disk outer radius $R_{\rm out}$ determines what portion of the NSC will interact with the AGN disk; thus it dictates how many BH will be embedded in the disk at the start of an AGN episode and how many NSC objects can be captured over the AGN disk lifetime.
Since opacity drives most of the processes affecting embedded BH, we use it to set $R_{\rm out}$.
The opacity of an SG disk roughly increases in the inner disk before dropping off significantly in the outskirts.
We set $R_{\rm out}$ to the radius at which the opacity decreases back to the value at the innermost disk radius, leading to a default disk size of $5\times10^4\,\Rg$, where $\Rg = GM/c^2$ is the gravitational radius (assuming a $10^{8}\Msun$ SMBH -- see Sec.~\ref{sec:smbh-mass}).
We adopt this value for the TQM disk since it also reaches a similar value as found at the innermost edge at $5\times10^4\,\Rg$, though the opacity behaves inversely with radius as compared to the SG model.
We expect larger disk sizes that intersect with a larger portion of the NSC to produce more mergers, although this depends on the NSC radial density profile.
Additionally, the initial pool of BH in a larger disk has greater distances through which to migrate and encounter possible companions that are unavailable in a smaller disk, possibly increasing the overall number of mergers.
As such, since larger objects migrate faster, they encounter more objects, so we might expect larger disk sizes to drive lower $q$ encounters.

\subsection{Orbital Eccentricity}
Gas drag can damp or pump orbital eccentricity $e$ of an embedded BH depending on the local conditions of the disk \citep{McK12, Secunda21, McKF24, SzolgyenMacLeodLoeb22}.
Radiation from accreting objects can cause an additional eccentricity pumping term \citep{Cornejo+23}, which we will test in future work. We compute the damping timescale according to \citet{Tanaka04} via \citet{McKF24} and apply \citet{Papaloizou00} exponential decay when $e<2H$ -- where $H$ is the pressure scale height -- and the characteristic decay timescale when $e>2H$ \citep{Bitsch10, Horn12} (see Paper I for details).
Once the orbital eccentricity is sufficiently small, BH migration and BBH formation become efficient.
A relaxed (thermal) NSC has median eccentricity of $e=0.7$, but this requires a long duty cycle between AGN phases.
SMBH spins observed in the local Universe are predominantly high spin with half $a>0.9$ \citep{Reynolds19}, indicating they preferentially accrete along a consistent plane rather than randomly \citep{Volonteri05}. If AGN consist of short-timescale accretion episodes \citep{Schawinski15,Shen21}, delivered from a fuel source in the same plane (possibly the dusty torus), this could drive consistent SMBH spin up over many AGN episodes. In this case, the NSC populations would have little time to relax between episodes, keeping eccentricities relatively damped (low).

We draw initial orbital eccentricities $e_0$ from a uniform distribution between 0 and $e_{0,max}$ representing the maximum eccentricity achieved during the prior quiescent state.
Thus, a smaller value of $e_{0,max}$ implies a shorter relaxation time between accretion episodes according to the previously mentioned scenario.
We choose fiducial values for the maximum in SG disks to be $e_{0,max}=0.5$ and in TQM disk to be $e_{0,max}=0.05$. The latter compensates for the lower density in TQM and correspondingly longer circularization timescales.
This choice assumes the NSC has insufficient time to fully relax between AGN disk episodes and that each such episode occurs in the same plane.
We also explore setting $e_0=0$ in the low density TQM model to bypass the long eccentricity damping time.

\subsection{Binary Formation}
We form BBH from the circularized BH population when the separation between adjacent orbits reaches one Hill radius ($\Rhill$) at the beginning of the time step, i.e. $\Delta a = |a_1-a_2| \leq 1\Rhill$, where $a$ is the semi-major axis, $\Rhill = a_1 \left( M_1/3M_{\bullet} \right)^{1/3}$, $M_1$ is the more massive candidate, and $\Mbul$ is the mass of the supermassive black hole. Recent analytic \citep{Qian+24, DodiciTremaine24} and hydrodynamical work \citep{DeLaurentiis+23, Whitehead+24} shows gas interactions complicate this process and allow BBH to form with a wide range of impact parameters and even between eccentric orbiters. Our simple model does not capture this rich complexity. We intend to implement models derived from these works in future versions of the code.

\subsection{Nuclear Star Cluster}
BH with disk-crossing orbits will align to and become fully embedded in the disk over time \citep{Fabj+20,Nasim22,Starfall22,Wang+24}. We use an NSC model with an outer radius of 5 pc and total mass $M_{\rm NSC}=3 \times 10^7\,\Msun$. The number of BH to stars is $N_{\rm BH}/N_{*} = 10^{-3}$ according to \citet{Generozov+18} and the typical mass ratio is $M_{\rm BH}/M_{*} = 0.1$. We draw locations from a broken power law distribution for the radial BH number density $n$ representing a ``cored'' type cluster where the radial number density index changes at a critical radius $R_{\rm crit}=0.25$ pc or $5\times10^4\,\Rg$ when $M_\bullet=10^8\,\Msun$. When $r<R_{\rm crit}$, $n\propto r^{-7/4}\,{\rm pc}^{-3}$ and $n\propto r^{-2.5}\,{\rm pc}^{-3}$ elsewhere \citep{BachallWolf76,Generozov+18}.

\subsection{SMBH Mass}
\label{sec:smbh-mass}
The mass of the central SMBH determines the profiles for a variety of important disk properties, including the density, scale height, and opacity, that vary with radius. It also influences the NSC mass and outer disk radius. For this study we restricted ourselves to a single SMBH mass of $10^8~\Msun$. We note that the standard TQM model uses an SMBH mass of $10^9~\Msun$, but Paper III finds $\sim 10^{8}\Msun$ is likely to be the dominant contributor to AGN channel merger rates. In future studies we will produce a realistic distribution of SMBH and NSC masses, appropriate to our Universe, and consider the impact of those parameters on the $q$--$\Xeff$ relationship.

\subsection{Retrograde Binary Fraction}
We choose the BBH angular momentum by drawing from a distribution biased by the user-chosen fraction $\fret$. BBH  are expected to form with prograde orbital angular momenta, i.e., aligned with the disk orbital angular momentum ($\vec{L}_{\rm bin}\,||\,\vec{L}_{\rm disk}$), because retrograde binaries may experience extreme eccentricity pumping \citep{Wang+22} or be flipped to prograde due to dynamical encounters \citep{Samsing22,McKF24}. For these reasons we choose $\fret=0$ as our default, but we  experiment with $\fret=0.1, 0.5$ to understand the impact of retrograde binaries on $q$--$\Xeff$ space. 


\subsection{Analysis Techniques}
Understanding how efficiently various AGN systems produce high $\Xeff$ events is crucial for understanding the underlying fraction of AGN events contributing to the observed population. We separate merger events into three groups based on the generations of the BH involved: (1) 1g mergers between first generation BH (1g-1g), (2) 2g mergers involving second generation BH (2g-1g and 2g-2g), and (3) $\geq$3g mergers involving a third generation BH or higher (3g-Ng). Sub-populations two and three are hierarchical mergers since they contain BH produced via prior merger events.

\citet{Callister+21} report an anti-correlation between $(q,\Xeff)$ with mean expectation values for the slopes of their ``observed" and ``predicted" samples of $d\chi/dq = -0.46$ and $-0.49$, respectively.
We measure the influence of hierarchical mergers using a non-linear least squares method to fit a line passing through the point $(q,\Xeff)=(1,0)$ to all mergers.
Shallower slopes (less negative) indicate a preference for larger $q$ and lower $\Xeff$ or simply a lack of high $\Xeff$ mergers. Steeper slopes (more negative) can indicate a few possibilities: many large $\Xeff$ low $q$ events, many large $\Xeff$ large $q$ events, or both.
We anchor our line of best fit at ($q=1,\Xeff=0$) where $q=1$ is the upper bound to $q=M_{2}/M_{1}$. 
In principle $\Xeff$ at $q=1$ is not bound.
However the weight of the numerically dominant 1g population consistently sits at $\Xeff=0$ (as shown below).
An unbound linear regression on the data in Fig.~\ref{fig:default-imf} decreases $d\chi/dq$ by $0.1$, and has a larger uncertainty of $\pm0.03$.
The value $R^2=0.12$ indicates a poor fit.
Thus, by fixing our line of best fit at $(q,\Xeff)=(1,0)$, we add a systematic uncertainty of $d\chi/dq=+0.1$, but obtain a better weighted fit overall.

\section{Results}
\label{sec:results}
Here we first explain the $q$--$\Xeff$ distribution of mergers from our fiducial models. Then, we demonstrate the effects of changing various input parameters on $q$--$\Xeff$ space.

\subsection{Reproducibility}
Data used in each panel of the figures in this paper can be reproduced using the correct seed and input parameters. In the \texttt{Makefile}, change the \texttt{SEED} parameter to the value listed in Table~\ref{tab:runs}. For random seeds using \texttt{Makefile}, add an empty line above \texttt{--seed}, comment out the \texttt{--seed} line and save the file. If a parameter needs to be changed, in each case, use the \texttt{model\_choice\_old.ini} file and change the appropriate parameter there.

\begin{deluxetable}{rcrc}
    \label{tab:run-parameters}
    \tabletypesize{\scriptsize}
    \tablewidth{0pt}
    \tablecaption{Unique parameters choices and random number generator seeds for each run. All other parameters are unchanged from those in the default models, which are otherwise identical to those in Paper I \citep{McKernan+24-McFACTS1}.}
    \tablehead{
        \colhead{Run} & \colhead{Seed} & \colhead{Parameter} & \colhead{Value}
    }
    \startdata
        \texttt{sg\_default} & 482821398 & \texttt{disk\_model\_name} & \texttt{sirko\_goodman} \\
         &  & \texttt{timestep\_num} & 50 \\
        \texttt{tqm\_default} & 783822028 & \texttt{disk\_model\_name} & \texttt{thompson\_etal} \\
         &  & \texttt{timestep\_num} & 300 \\
         &  & \texttt{disk\_radius\_trap} & 500 \\
         &  & \texttt{disk\_bh\_orb\_ecc\_max\_init} & 0.05 \\
        \hline
        \texttt{sg\_g1} & 40310999 & \texttt{nsc\_imf\_powerlaw\_index} & 1 \\
        \texttt{tqm\_g1} & 63434937 & \texttt{nsc\_imf\_powerlaw\_index} & 1 \\
        \hline
        \texttt{sg\_t0p25} & 27228638 & \texttt{timestep\_num} & 25 \\
        \texttt{sg\_t0p75} & 46249919 & \texttt{timestep\_num} & 75 \\
        \texttt{tqm\_t2p5} & 75904298 & \texttt{timestep\_num} & 250 \\
        \texttt{tqm\_t5} & 43913681 & \texttt{timestep\_num} & 500 \\
        \hline
        \texttt{sg\_r2e4} & 61693043 & \texttt{disk\_radius\_outer} & $2\times10^4$ \\
        \texttt{sg\_r7e4} & 572938 & \texttt{disk\_radius\_outer} & $7\times10^4$ \\
        \texttt{tqm\_r2e4} & 37960990 & \texttt{disk\_radius\_outer} & $2\times10^4$ \\
        \texttt{tqm\_r7e4} & 2039485 & \texttt{disk\_radius\_outer} & $7\times10^4$ \\
        \hline
        \texttt{sg\_s0p02} & 13555771 & \texttt{nsc\_bh\_spin\_dist\_sigma} & 0.02 \\
        \texttt{sg\_s0p20} & 33286099 & \texttt{nsc\_bh\_spin\_dist\_sigma} & 0.20 \\
        \texttt{tqm\_s0p02} & 74447390 & \texttt{nsc\_bh\_spin\_dist\_sigma} & 0.02 \\
        \texttt{tqm\_s0p20} & 10409162 & \texttt{nsc\_bh\_spin\_dist\_sigma} & 0.20 \\
        \hline
        \texttt{sg\_fr0p1} & 38049133 & \texttt{fraction\_bin\_retro} & 0.1 \\
        \texttt{sg\_fr0p5} & 28559955 & \texttt{fraction\_bin\_retro} & 0.5 \\
        \texttt{tqm\_fr0p1} & 10171242 & \texttt{fraction\_bin\_retro} & 0.1 \\
        \texttt{tqm\_fr0p5} & 59253870 & \texttt{fraction\_bin\_retro} & 0.5 \\
        \hline
        \texttt{sg\_e0p1} & 60447293 & \texttt{disk\_bh\_orb\_ecc\_max\_init} & 0.1 \\
        \texttt{sg\_e0p5} & 33885262 & \texttt{disk\_bh\_orb\_ecc\_max\_init} & 0.5 \\
        \hline
        \texttt{sg\_e0p7} & 36123383 & \texttt{disk\_bh\_orb\_ecc\_max\_init} & 0.7 \\
        \texttt{sg\_e0p7\_t0p75} & 48962233 & \texttt{disk\_bh\_orb\_ecc\_max\_init} & 0.7 \\
         &  & \texttt{timestep\_num} & 150 \\
        \hline
        \texttt{tqm\_e0p015} & 85532667 & \texttt{disk\_bh\_orb\_ecc\_max\_init} & 0.015 \\
        \texttt{tqm\_e0p1} & 20803120 & \texttt{disk\_bh\_orb\_ecc\_max\_init} & 0.1 \\
        \hline
        \texttt{tqm\_e0\_t1} & 60471184 & \texttt{disk\_bh\_orb\_ecc\_max\_init} & 0.0 \\
         &  & \texttt{timestep\_num} & 100 \\
        \texttt{tqm\_e0} & 92269440 & \texttt{disk\_bh\_orb\_ecc\_max\_init} & 0.0
    \enddata
    \tablecomments{The default timestep is set to $10^4$ yrs. Set the corresponding number of time steps with \texttt{timestep\_num} to achieve the desired disk lifetime. Set migration trap locations using \texttt{disk\_radius\_trap} in $\Rg$. \texttt{disk\_bh\_orb\_ecc\_max\_init} sets the maximum initial orbital eccentricity allowed for orbits of BH initially embedded in the disk. Use \texttt{nsc\_imf\_powerlaw\_index} to control the slope of the probability distribution function used for mass draws. \texttt{disk\_radius\_outer} controls the location of the outer edge of the disk in $\Rg$. \texttt{nsc\_bh\_spin\_dist\_sigma} controls the width of the normal distribution from which initial BH spins are drawn. \texttt{fraction\_bin\_retro} controls the fraction of retrograde binaries allowed to form with angular momenta pointing opposite to the disk's.}
\end{deluxetable}

\subsection{Default Models}
\begin{figure*}
    \centering
    \includegraphics{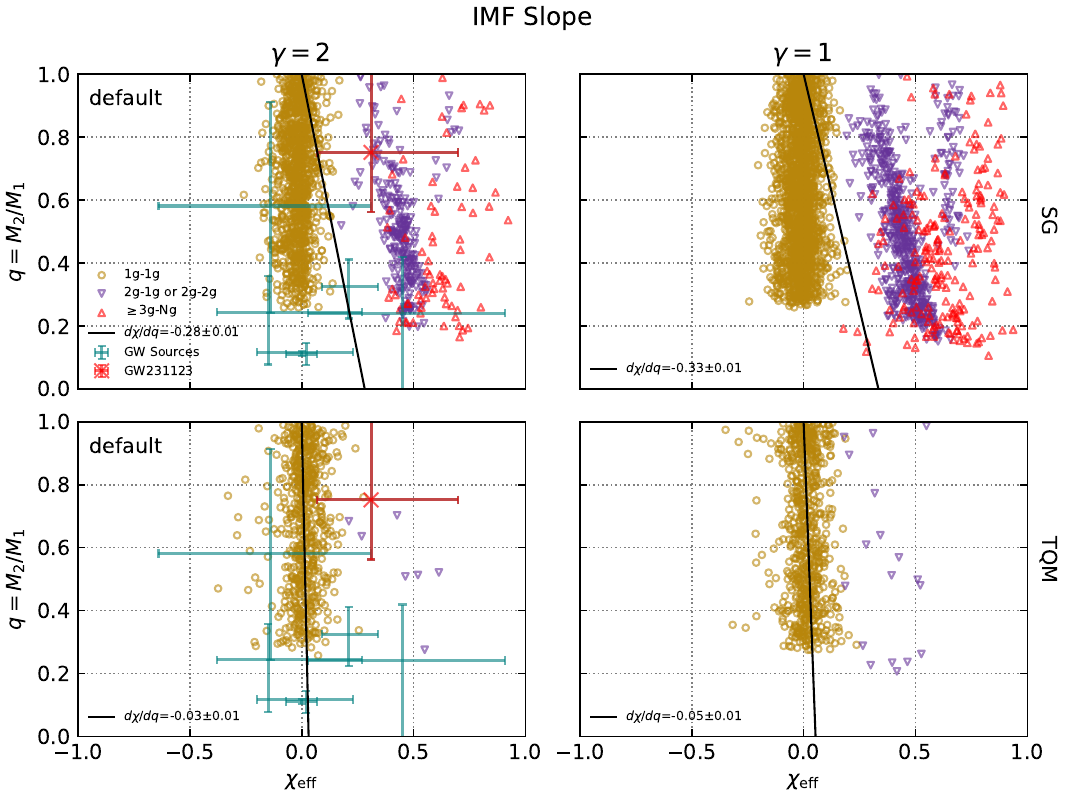}
    \caption{\textbf{Mass ratio ($q$) distribution as a function of $\Xeff$}.
    \textit{Top left:} results from our default setup for a \citet{SirkoGoodman03} disk with BH mass distribution power-law index $\gamma=2$, run \texttt{sg\_default}.
    Gold points represent 1g-1g mergers (see text).
    Inverted purple triangles indicate either 2g-1g or 2g-2g mergers. Red triangles indicate $\geq$3g-Ng mergers.
    The solid and dashed lines are fit through ($\Xeff,q)=(0,1)$ and entire and $\geq$2g populations, respectively.
    Teal points with error bars are a representative sample of GW sources chosen to avoid cluttering and provide context for our results: GW190412, GW190521, GW190814, GW200115\_042309, GW200208\_222617, GW200210\_092254.
    The red orange data point with an ``x" represents GW231123.
    \textit{Bottom left:} default setup for a \citet{Thompson+05} disk, run \texttt{tqm\_default}.
    \textit{Right column:} same as SG and TQM defaults but with shallower IMF index of $\gamma=1$ that increases the probability for high-mass BH. Each experiment was performed for 100 galaxies.
    See Table~\ref{tab:run-parameters} for setup parameters and Table~\ref{tab:popstats} for population statistics.}
    \label{fig:default-imf}
\end{figure*}

The top left panel of \Fig{fig:default-imf} shows the mass ratio ($q$) versus effective spin ($\Xeff$) phase space for our default model consisting of 100 galaxies using an SG disk with lifetime $\tauAGN = 0.5$ Myr, radius $R_{\rm out} = 5\times 10^4\,\Rg$, and migration trap at 700$\,\Rg$. We refer to this model as \texttt{sg\_default} and set the quantities using these \texttt{McFACTS} parameters: \texttt{timestep\_num}, \texttt{disk\_radius\_outer}, and \texttt{disk\_radius\_trap}. The initial BH population is drawn from a mass distribution of index $\gamma=2$ (\texttt{nsc\_imf\_powerlaw\_index}) with a maximum initial orbital eccentricity $e_{\rm 0,max}=0.5$ (\texttt{disk\_bh\_orb\_ecc\_max\_init}). We only allow prograde BH orbiters to form BBH and do not allow the retrograde orbiters to interact in any way. We restrict BBH to form from BH orbiting in a prograde orientation with the same sense of direction as the disk ($\vec{L}_{\rm BH}\,||\,\vec{L}_{\rm disk}$) and bias the resulting binaries towards a prograde direction (i.e., $\vec{L}_{\rm bin}\,||\,\vec{L}_{\rm disk}$, and see an exploration of $\fret$ below). Retrograde BH orbiters do exist but presently are not allowed to form BBH or interact with the prograde BH population, although they can (and do) flip to prograde over time (see Paper I).

Gold points correspond to mergers among the first BH generation (1g-1g; hereafter 1g mergers). Inverted purple triangles are mergers involving at least one 2g BH, i.e. either 2g-1g or 2g-2g (hereafter 2g mergers). Red triangles indicate mergers involving at least one 3g or higher BH ($\geq$3g-Ng; hereafter $\geq$3g mergers). The mean and standard deviation among the gold, purple, and red distributions, respectively, are $(\barq,\barXeff)=\{[0.67\pm 0.19,\ -0.02\pm 0.05], [0.51\pm 0.18,\ 0.44\pm 0.08], [0.45\pm 0.22,\ 0.61\pm 0.13]\}$ (see also Table~\ref{tab:popstats}).

The points lie in roughly two distributions: 1g mergers stay clustered around $\Xeff \sim 0$, whereas 2g and $\geq$3g tend to have larger, positive $\Xeff$ values and trend towards smaller $q$ values. Since $10\,M_{\odot}$ BH are the most common from the IMF, then those in the 2g population will tend to be $\sim 20\,M_{\odot}$ leading to 2g-1g mergers concentrating around $q \sim 0.5$. Similarly, a given 3g-1g merger will typically concentrate at $q \sim 1/3$.

We quantify the anti-correlation between $\Xeff$ and $q$ (larger $\Xeff$ values are measured for lower $q$) using the slope of a line fit to the data and passing through $(q,\Xeff)=(1,0)$. In the top left panel of \Fig{fig:default-imf} the black solid line (fit to all points) has a slope of $d\chi/dq=-0.28\pm0.01$.

The bottom left panel of \Fig{fig:default-imf} shows the results from our default TQM setup ($\tqmdefault$). All parameters are identical to those in the $\sgdefault$ setup except the TQM disk model, a longer lifetime at $\tauAGN=3$ Myr, and lower maximum initial orbital eccentricity $e_{\rm 0,max}=0.05$. The 1g population is qualitatively similar as those occurring in the SG disk and concentrate around $\Xeff=0$, though spread over a wider range of $\Xeff$, and the hierarchical mergers still shift towards positive $\Xeff$, but only seven 2g events and zero $\geq3g$ events occurred compared to the 606 1g mergers. The low density TQM disk produces drastically less mergers than a more dense SG disk despite a lifetime three times longer. Additionally, low disk opacity results in radiative feedback torques that dominate over inward migration torques causing a net outward migration for more massive objects that increases time between encounters.

The mean mass ratio $\barq=0.68\pm0.20$ for 1g mergers (gold points) in the TQM default run is similar to the SG default. The spins are centered on zero but with larger standard deviation, $\barXeff({\rm 1g})=-0.01\pm-0.07$. The wider spread in $\Xeff$ in TQM comes from a combination of (i) a lower surface density, leading to a longer damping time (allowing more eccentric 1g BH to spin down to more negative $\Xeff$ for longer) as well as (ii) a longer overall disk lifetime (3 Myr) allowing more circularized 1g BH to spin up for longer. 

The 2g sample (purple triangles) has a mean mass ratio of $\barq({\rm 2g})=0.55\pm0.14$ and $\barXeff({\rm 2g})=0.44\pm0.14$, though these statistics are for only seven events. The paucity of hierarchical mergers results in an essentially flat $d\chi/dq=-0.03\pm0.01$ slope.

Qualitatively, both plots tell us populations of higher generation mergers tend to have higher \textit{positive} $\Xeff$. This trend is visible in $\overline{\Xeff}$ (see Table~\ref{tab:popstats}), reflecting what was seen in \citet[][see Fig. 2]{Mckernan+22:q-Xeff}. Because we initialized BH spin orientations randomly, we expect $\Xeff\sim0$ for 1g events. However, their remnants receive a natal spin of $\chi \sim 0.7$ directed along the progenitor binary's angular momentum $\vec{L}_{\rm bin}$. Since we limit BBH to form prograde by default, $\vec{\chi}$ tends to point parallel to the disk angular momentum (positive). A BH's mass and spin magnitude increase with generation, so we expect the hierarchical component of BBH mergers to be both more massive and have a higher spin, dominating \eq{eqn:Xeff} and leading low $q$ events to have high $\Xeff$ values matching our findings (but see \sect{sec:retro-results} for what happens when some fraction $\fret$ of BBH form retrograde).

We include a selection of GW sources in the left column of Fig.~\ref{fig:default-imf} as teal data points: GW190412, GW190521, GW190814, GW200115\_042309, GW200208\_222617, GW200210\_092254, 
and finally the location of the recent event  GW231123, is marked with a red-orange ``x"
\citep{GW190521astro,GW190521,Abbot+2023:O3b,Abbot+24:GWTC-2.1,LVK+25-GW231123}.
We use reported mass ranges to calculate the estimated errors in the mass ratio. The central point near $q=0.6$ is GW190521 and overlaps mostly with the 1g population. 
However, the error in $\Xeff$ is great enough for this source to lie in the 2g-2g population of hierarchical mergers, supporting the claim that the progenitors reached masses $98.4_{-21.7}^{+33.6}\,\Msun$ and $57.2_{-30.1}^{+27.1}\,\Msun$ via prior mergers . Alternatively, if $\chi_{\rm eff}<0$ for GW190521, this might indicate an origin in AGN with $f_{\rm ret}>0$ (see discussion below). 
The point at $\Xeff=-0.15$ and $q=0.24$ (masses $5.9_{-2.5}^{+2.0}\,\Msun$ and $1.44_{+0.85}^{-0.28}\,\Msun$) is GW200115\_043209.
The progenitors of this source have masses below the minimum mass in our IMF, but the mass ratio and effective spin suggest this source belongs to the 1g-1g population.
GW190814 and GW200210\_092254, both with $q=0.1$ and centered at $\Xeff=0$, also correlate with the 1g-1g merger distribution. 
Our IMF range limits the mass ratio range of 1g-1g mergers to $q>$0.25, but see  Paper III for a study producing mergers down to $q \sim 0.1$. GW190412 falls between the 1g and hierarchical populations, but their masses ($27.7_{-6.0}^{+6.0}\,\Msun$ and $9.0_{-1.4}^{+2.0}\,\Msun$) make it difficult to distinguish between these possible progenitor populations. 
GW200208\_222616 lies at  $\Xeff=0.45$ and $q=0.24$ (masses $51_{-30}^{+103}\,\Msun$ and $12.3_{-5.5}^{+9.2}\,\Msun$).
Though the data point overlaps with the low-$q$ end of the hierarchical population we produced, the large error bars, especially on the primary mass, suggest a 1g-1g or hierarchical merger are both possible explanations.
Finally, the red-orange ``x" marks GW231123 \citep{LVK+25-GW231123} which lies directly in the 2g population island.
GW231123 overlaps with the high-$\Xeff$ and high-$q$ region primarily populated by $\geq$3g events that results from components with large, aligned spins of similar masses.
This suggests that GW231123 is a hierarchical merger between progenitors that previously merged in a similar plane (see Delfavero et al. 2025 in prep.)
 
\begin{figure*}
    \centering
    \includegraphics{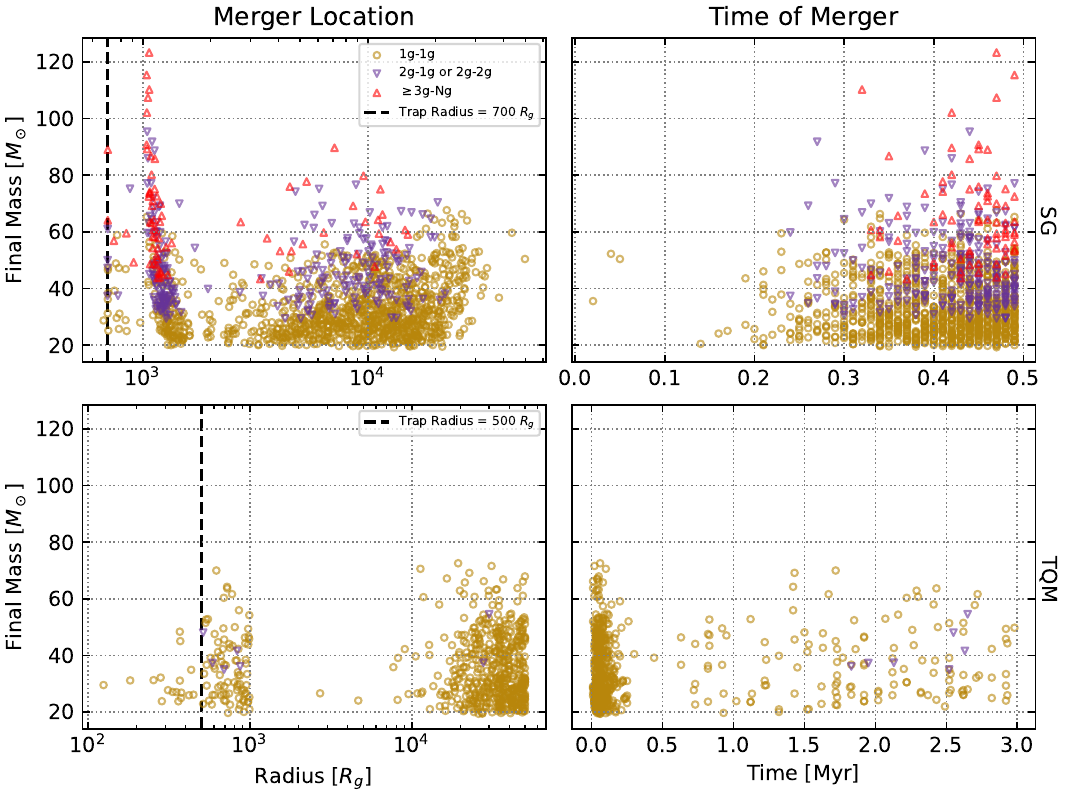}
    \caption{Final mass of merger remnants versus radius (left column) and time (right column) in SG disks (top row) and TQM disks (bottom row).
    Strong migration torques and traps in SG disks near $10^3\,\Rg$ accelerate hierarchical mergers whereas weaker torques in the TQM disk cause traps to be less important leading to a maximum of 3g remnants and lower masses.
    SG disks take $\sim0.2$ Myr to regularly produce 1g mergers with 2g mergers starting at 0.25 Myr and 3g+ mergers at 0.33 Myr, while final mass increases steadily with time and reaches $\sim 120\,\Msun$.
    TQM disks quickly merge the initial embedded BH population in the first 0.25 Myr and reach a steady state as the disk captures more BH.
    The mass range of remnants stays relatively even between 20 and 70 $\Msun$ throughout time.
    Markers are as in \Fig{fig:default-imf}.}
    \label{fig:default-radius-time}
\end{figure*}

We show the final masses as a function of radius in the left column and time in the right column in  \Fig{fig:default-radius-time}.
In an SG disk (top row), mergers between all generations occur beyond $3\times10^3\,\Rg$, though 1g-1g mergers dominate this region.
As migration torques drive the embedded population inwards, a reduced torque at $\lesssim 10^{3}r_{g}$ causes a traffic jam as they approach the trap set at $\sim 700r_{g}$ \citet{Bellovary16} (also the vertical dashed line). For our relatively short-lived default $\sim 0.5$Myr disk, most hierarchical mergers occur in this 'swamp' outside of the trap (see also discussion in Paper I). The dense inner disk population here as a result of migration and disk capture, drives the majority of $\geq$2g mergers and produces all remnants with mass $>90\,\Msun$, reinforcing the importance of traps in SG disks for producing BH with masses in the upper mass gap.
The circularization, migration, binary formation, and binary hardening processes take roughly $\sim 0.2$ Myr at which point 1g mergers occur,
consistent with realignment timescales of inclined binaries at migration traps  \citep[see Fig. 2 of][]{Gilbaum+25}.
After $\sim 0.25$Myr, 2g mergers appear, and $\geq$3g mergers start around $\sim 0.33$ Myr.
In each population, the final mass attainable increases with time.

In a TQM disk (bottom row of \Fig{fig:default-radius-time}) torques from radiative feedback combined with low disk opacity cause outward migration towards the disk edge at $5\times10^4\,\Rg$, while inside $10^3\,\Rg$, the torques are more balanced, leading to mergers down to $10^2\,\Rg$.
The TQM disk produces a burst of mergers in the first 0.25 Myr from BH on initially nearly circular orbits.
Then, the rate settles to a steady-state sustained by BH captured from the NSC by the disk.
The much weaker migration torques in TQM disks are unable to concentrate BH as densely as SG disks near migration traps suppressing hierarchical mergers to 3g remnants up to $\sim70\,\Msun$.

\begin{figure}
    \centering
    \includegraphics[width=1.0\columnwidth]{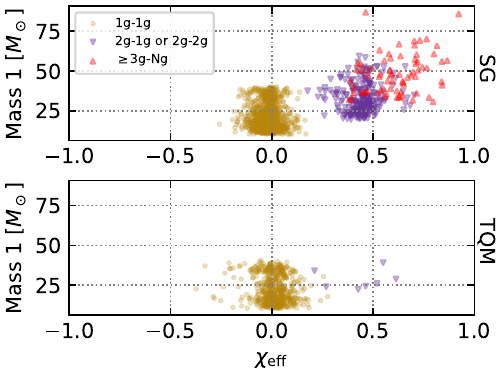}
    \caption{\textbf{Primary Mass vs. $\Xeff$.}
    Hierarchical mergers drive primary masses higher and produce mergers with larger values of $\Xeff$.
    Markers are as in \Fig{fig:default-imf}.}
    \label{fig:default-primarymass-xeff}
\end{figure}

Fig.~\ref{fig:default-primarymass-xeff} shows the primary mass of mergers as a function $\Xeff$.
Hierarchical mergers drive up the primary mass of subsequent events which tend to have larger $\Xeff$.
In high generation mergers whose primaries are more likely to have higher mass, randomly oriented spins of companions have less effect on $\Xeff$ since the primaries are remnants that have inherited their spin from the orbital angular momentum of their progenitor BBH, and we have restricted BBH to form with angular momenta aligned with the disk in this setup.

Below we explore the effects of a series of parameters on the intrinsic $q$--$\Xeff$ distributions produced by galaxies hosting a SG or TQM AGN disk model.

\subsection{Changing the Mass Function}
First, we test the effect of flattening the BH initial mass function (IMF) by changing the power-law index from a steeper $\gamma=2$ (our default) to $\gamma=1$ (\texttt{nsc\_imf\_powerlaw\_index} in \texttt{McFACTS}). NSCs may have a shallower-than-Salpeter IMF, and indeed a shallow IMF is observed in our own Galactic nucleus \citep[][]{Lu+2013}.

The top right panel of \Fig{fig:default-imf} shows the results from run \texttt{sg\_g1} and is qualitatively similar to $\sgdefault$. We see the same population islands: a 1g merger population centered on $\Xeff \sim 0$, as well as hierarchical populations at larger $\Xeff$. However, given a shallower mass function or equivalently a heavier IMF, the typical mass ratio shifts to smaller values, as expected. Thus, among the 1g-1g population we can see that the population is dispersed away from $q \sim 1$, evening out the slight tapering seen in the top left panel were $\gamma=2$. From Table~\ref{tab:popstats}, among 1g mergers $\barq$ drops slightly from $\barq$(1g)$=0.67\pm0.19$ from \texttt{sg\_default} to $\barq$(1g)$=0.63\pm0.21$ in \texttt{sg\_g1} with double the number of mergers, while $\barXeff$(1g) remains the same ($-0.02\pm0.05$ vs. $-0.01\pm0.06$ here). Among higher generation mergers, visually there is a slight smearing out in both $q,\Xeff$ with the large $q$ region filling in. Although $\barq,\barXeff$ remain similar to the default within the margins of error for both hierarchical populations, the shallower IMF produces double the number of hierarchical mergers overall and a factor three more mergers of $\geq$3g (see Table~\ref{tab:popstats}).

The slope of the line measuring the anti-correlation decreases significantly from $d\chi/dq=-0.28\pm0.01$ to $d\chi/dq=-0.33\pm0.01$ with shallower IMF.
The change in slope seems mainly driven by the ability for more hierarchical mergers to form, given the larger Hill radius of heavier BH, which increases the likelihood of BBH formation.

As we flatten the BH IMF from $\gamma=2$ in the $\tqmdefault$ setup to $\gamma=1$ for setup \texttt{tqm\_g1}, we do not see any visual difference comparing the two panels in the bottom row in \Fig{fig:default-imf}. The 1g population shifts slightly down to $\barq({\rm 1g})=0.65\pm0.23$ while $\barXeff$ is identical despite 10\% more mergers. The 2g population keeps the same $\barq$(2g) but doubling the number of events and increasing the number of low $q$ mergers drives $\barXeff$(2g) to $0.37\pm0.12$. As for $\gamma=2$, zero $\geq$3g mergers occurred. The line fit slope does not significantly change between $\tqmdefault$ and \texttt{tqm\_g1}.

For denser disks like SG, hierarchical mergers occur more readily as we flatten the mass function, since migration speed depends linearly on the mass (see Paper I). The general increase in number of large $\Xeff$ hierarchical events of $\geq$3g can most easily be seen in the steepening of the line fit to the overall population. Notably, the only appreciable change occurs for $q$ of 2g mergers when changing IMF slope in low density, low torque disks like TQM.

\subsection{Accretion Disk Lifetime}
Next, we explore changes in the disk lifetime from our default models. We consider SG disk lifetime spanning $\tauAGN=[0.25,0.75]$ Myr. The less dense TQM disk model spans $\tauAGN=[2.5,5]$ Myr. These adjustments are made by changing \texttt{timestep\_num} according to \texttt{timestep\_duration\_yr} (default $10^4\,$yr).

\begin{figure*}
    \centering
    \includegraphics{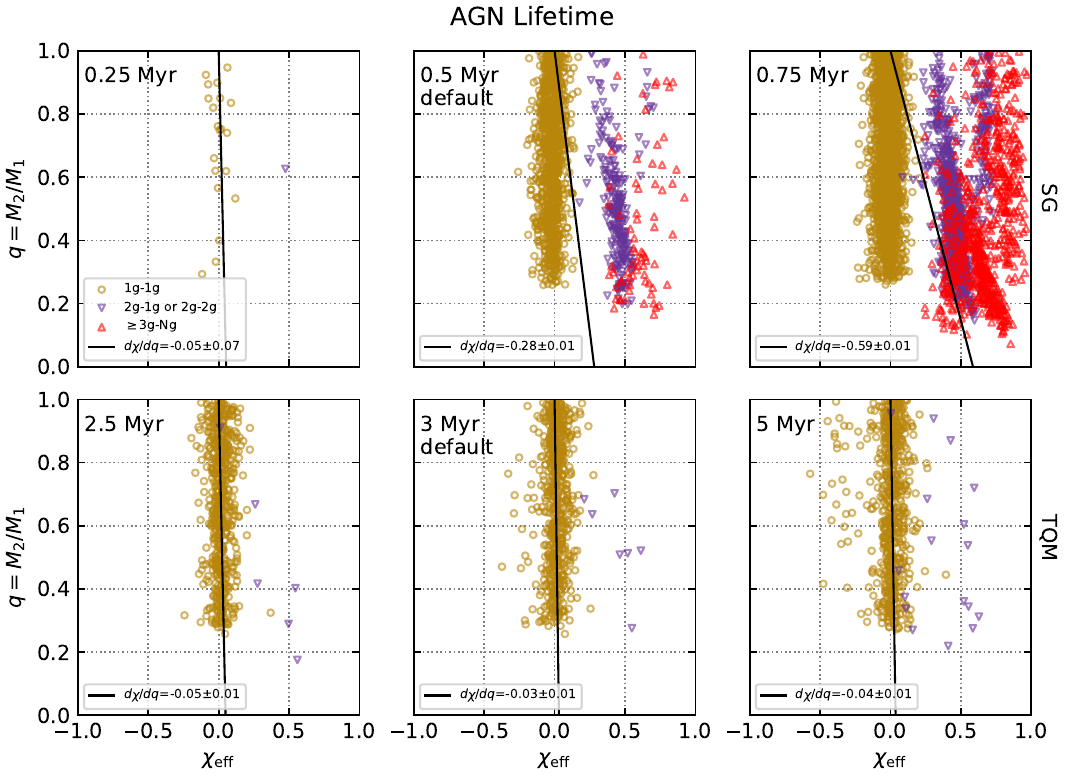}
    \caption{\textbf{Varying the disk lifetime.}
    \textit{Top row:} SG disk setups with $\tauAGN=[0.25,0.5,0.75]$ Myr.
    \textit{Bottom row:} TQM disk setups with $\tauAGN=2.5$, 3, and 5 Myr. The anti-correlation steepens with lifetime in dense SG disks as more hierarchical mergers with large $\Xeff$ occur, but $\tauAGN$ has little impact on low density TQM disks.
    Markers are as in \Fig{fig:default-imf}.
    See Table~\ref{tab:run-parameters} for unique setup parameters and Table~\ref{tab:popstats} for population statistics.}
    \label{fig:time}
\end{figure*}

The top row of \Fig{fig:time} shows the results for a SG disk with $\tauAGN=0.25\,$Myr (left, \texttt{sg\_t0p25}) 0.5 Myr (center, $\sgdefault$), and 0.75 Myr (right, \texttt{sg\_t0p75}). Decreasing $\tauAGN$ decreases the number of 1g and produces a single 2g merger since few BH on eccentric orbits are able to circularize, migrate, form binaries, and merge in such a short $\tauAGN$. Increasing $\tauAGN$ does not significantly change the population features, however the increased time allows $\geq$3g mergers to fill in the low $q$ and high $\Xeff$ (bottom right) corner of the space. The island in the upper right region is populated by mergers of BBH where both components are high generations and of similar mass and therefore have spins near the natal value of $\Xeff \sim 0.7$ for merger remnants.

The bottom row of \Fig{fig:time} shows the results for a TQM disk with $\tauAGN=2.5$ Myr (left, \texttt{tqm\_t2p5}) 3 Myr (center, $\tqmdefault$), and 5 Myr (right, \texttt{tqm\_t5}). Both 1g and 2g populations look very similar except for a slight increase in the number of 2g mergers and some 1g mergers with more negative $\Xeff$ when $\tauAGN=0.75$ Myr. This latter feature is caused by eccentric BH drastically spinning down during the long circularization process required for this model, but there are no statistically significant changes to any of the diagnostic quantities that are not attributable to low numbers (see Table.~\ref{tab:popstats}). 

In SG disks, $\tauAGN$ greatly affects the slope of the anti-correlation even when we exclude the case when $\tauAGN=0.25$ Myr due to the lack of mergers.
Increasing $\tauAGN$ from 0.5 Myr to 0.75 Myr moves fit from $d\chi/dq=-0.28\pm0.01$ to $-0.59\pm0.01$, caused by five times more 2g mergers and ten times more $\geq$3g mergers.
This overall steepening primarily points to a correlation between $\tauAGN$ and the number of hierarchical mergers, especially between high generation progenitors.

\begin{figure*}
    \centering
    \includegraphics{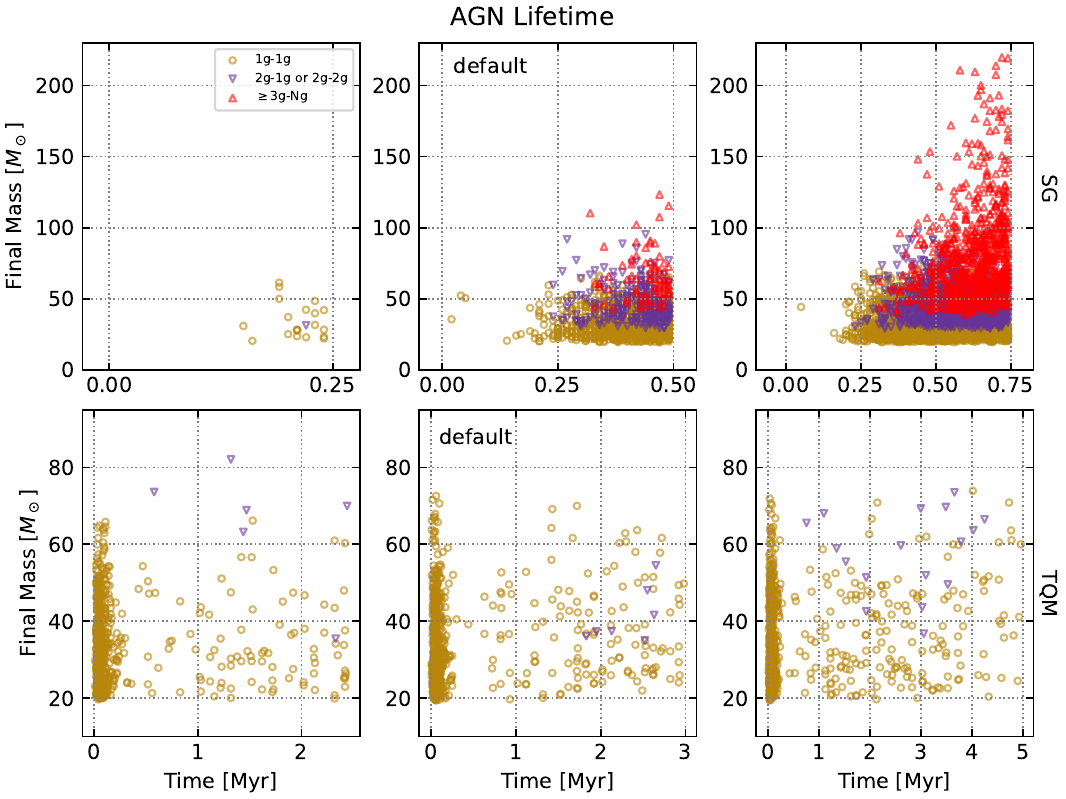}
    \caption{\textbf{Merger remnant mass versus disk lifetime.}
    \textit{Top row:} SG disk setups with $\tauAGN=$ 0.25, 0.5, and 0.75 Myr.
    \textit{Bottom row:} TQM disk setups with $\tauAGN=2.5$, 3, and 5 Myr.
    Hierarchical mergers drive up the mass of remnants in SG disks with increased lifetime, whereas TQM disks quickly merge BH on near circular orbits but low migration rates suppresses subsequent interactions to increase remnant masses beyond $\sim80\,\Msun$.
    Markers are as in \Fig{fig:default-imf}.
    See Table~\ref{tab:run-parameters} for unique setup parameters and Table~\ref{tab:popstats} for population statistics.}
    \label{fig:time-mass}
\end{figure*}

Fig.~\ref{fig:time-mass} shows the history of mergers for the runs shown in Fig.~\ref{fig:time}.
As $\tauAGN$ increases, an SG disk steadily produces higher masses via hierarchical mergers.
The $\geq$3g population builds masses up to 100$\,\Msun$ in 0.5 Myr and 200$\,\Msun$ after 0.75 Myr due to saturation at the migration traps.
In contrast, a TQM disk does not produce increasingly massive remnants via hierarchical mergers with longer lifetime.
Not shown here, the behavior in TQM disks continues as long as 10 Myr.

\subsection{Accretion Disk Radius}
\begin{figure*}
    \centering
    \includegraphics{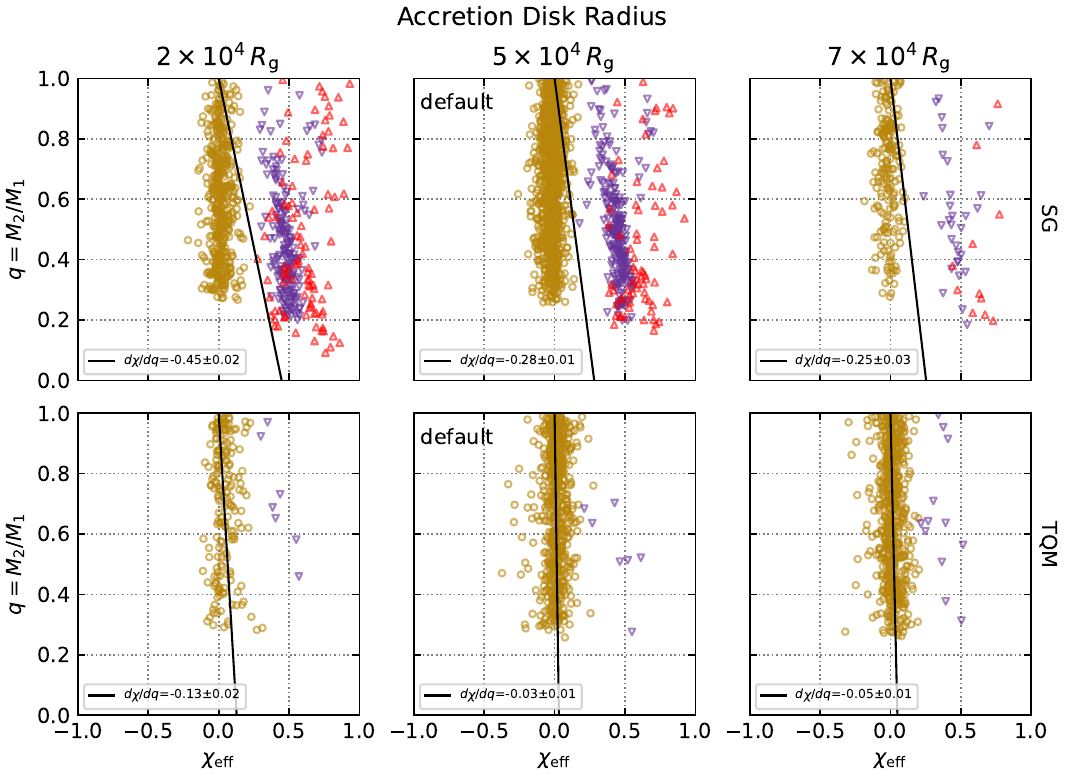}
    \caption{\textbf{Varying disk radius.}
    SG and TQM disks with radii $r = (2\times 10^{4}, 5\times 10^{4},\,10^{5})\ R_{\rm g}$.
    Both decreases and increases in disk radius result in fewer mergers. Model name left-to-right, top-to-bottom: \texttt{sg\_r2e4}, \texttt{sg\_default}, \texttt{sg\_r7e4}, \texttt{tqm\_r2e4}, \texttt{tqm\_default}, and \texttt{tqm\_r7e4}.
    Markers are as in \Fig{fig:default-imf}.
    See Table~\ref{tab:run-parameters} for unique setup parameters and Table~\ref{tab:popstats} for population statistics.}
    \label{fig:radius}
\end{figure*}

Next, we vary the radial extent of each disk model to $R_{\rm out}=2\times10^4$ and $7\times10^4$ $\Rg$ by changing \texttt{disk\_radius\_outer}, which affects the initial number of BH in the disk. \Fig{fig:radius} shows the relevant results.

Smaller disks have a smaller cross section with the NSC and will naturally result in a reduction in the BH merger rate. This is reflected in the SG setup with $R_{\rm out} = 2\times10^4\,\Rg$ (top left, \texttt{sg\_r2e4}). Note the number of mergers \emph{also} decreases with larger disk size (top right, \texttt{sg\_r7e4}). In a slightly larger disk, we do not add significantly more BH. This is because our default BH radial density function has a powerlaw break at $R=5 \times 10^{4}r_{g}$, with a steeper power-law index outside ($-2.5$ compared to inside $-7/4$). As a result, even though the disk is larger, we add proportionally fewer BH and so the average spacing of BH over the range $[\texttt{disk\_radius\_inner, disk\_radius\_outer}]$ must \emph{increase} in our model, thus driving down the average rate.

The line fit for mergers produced in a smaller disk (\texttt{sg\_r2e4}) \emph{increases} to $d\chi/dq=-0.45\pm0.02$ due to a more equal number of mergers in the 1g and $\geq$2g populations as compared to \sgdefault. A larger disk in \texttt{sg\_r7e4} disk produces a similar slope to \sgdefault\ since the number of events in each population decreased proportionally.

The already low number of mergers in the TQM disk goes decreases further with a smaller disk radius (bottom left, \texttt{tqm\_r2e4}). While the SG model saw a \emph{decrease} in mergers for larger disks, the TQM model sees a constant number of mergers when $R_{\rm out}=7\times10^4\Rg$ (bottom right, \texttt{tqm\_r7e4}, due to more room for out-migrating BH due to thermal torques in the low opacity outer disk.

Statistical measures for 1g mergers in the TQM models do not significantly change, but the low number of hierarchical mergers causes ($\barq$(1g), $\barXeff$(1g)) to fluctuate from ($0.71\pm0.17, 0.43\pm0.09$) in \texttt{tqm\_r2e4} to ($0.55\pm0.14, 0.44\pm0.08$) in \tqmdefault\ and ($0.65\pm0.20, 0.36\pm0.09$) in \texttt{tqm\_r7e4}. This has a strong influence on the measured slope which decrease as radius increases $d\chi/dq=[-0.13\pm0.02, -0.03\pm0.01, -0.05\pm0.01]$.

Note: see Paper III for the effect on the merger rate for self-consistently changing outer disk radius with SMBH mass.

\begin{figure*}
    \centering
    \includegraphics{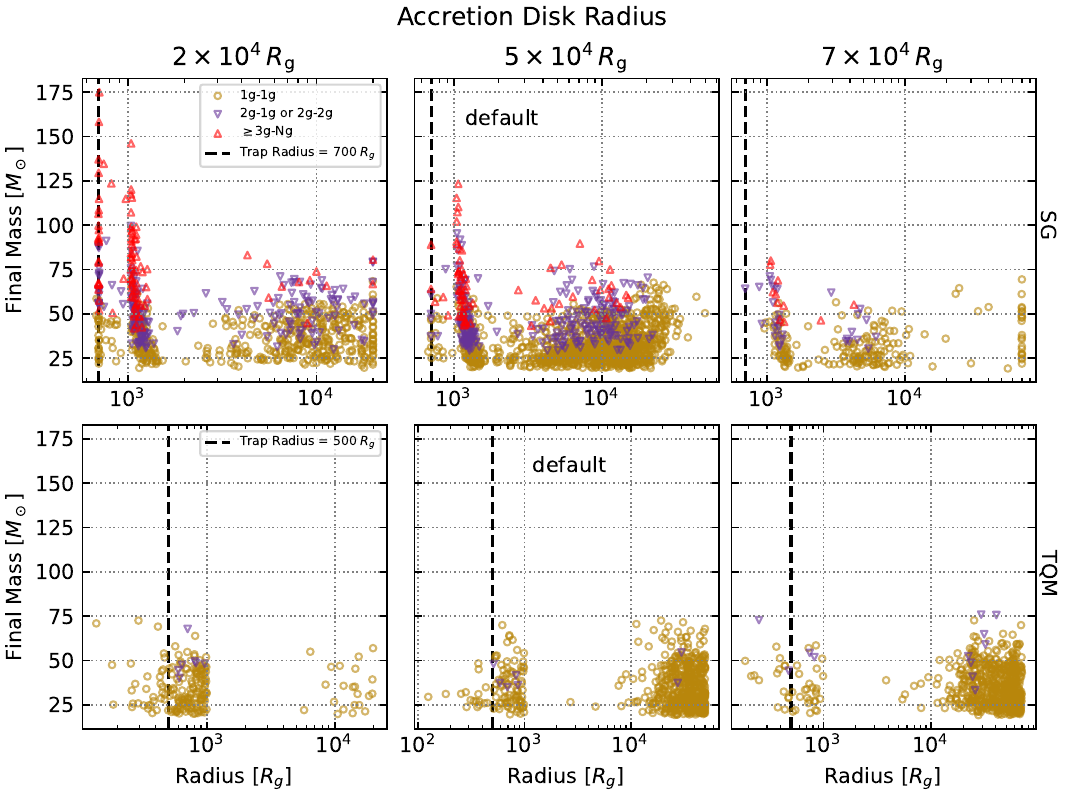}
    \caption{\textbf{Merger remnant mass versus radius.} Left column shows outer disk radii set to $2\times10^4\,\Rg$ compared to the default $5\times10^4\,\Rg$ in the center column, and $7\times10^4\,\Rg$ in the right column.
    SG disks produces less mergers with both changes for SG (top row).
    Shrinking a TQM (bottom row) disk also produces less mergers, but extending this disk includes more of the optically thick outskirts that drives migration in this model so the number of mergers is the same as the default case.
    Markers are as in \Fig{fig:default-imf}.}
    \label{fig:radius-mass}
\end{figure*}

Fig.~\ref{fig:radius-mass} shows final masses of the mergers as a function of merger location for the corresponding runs shown in Fig.~\ref{fig:radius}.
The trap at $700\,\Rg$ and migration slowdown at $10^3\,\Rg$ produce the majority of hierarchical mergers and high mass remnants in the SG disk, while the slow migration rate in the TQM disk reduces the traps' effectiveness.

For an SG disk, decreasing the radius to $2\times10^4\,\Rg$ produces roughly half the number of mergers (750) versus the default (1412) with radius set to $5\times10^4\,\Rg$, while increasing the radius to $7\times10^4\,\Rg$ significantly reduces the number (214). As outlined above, average initial separations between BH increase, suppressing e.g. the importance of traps for a default  $\tauAGN = 0.5$ Myr. An increase to $\tauAGN =0.6$ Myr, yields more time to migrate, producing 489 mergers and the hierarchical population reemerges near the trap.

Changing the disk size in a TQM model produces slightly different behavior for our choices of radii.
Like SG disk, smaller TQM disks overlap with less of the NSC, but the low inner disk opacity reduces the effectiveness of migration and traps and produces only 194 mergers relative to the default size that produces 615 mergers.
By contrast, the higher opacity at large radii in TQM disks drives migration to compensate for the effect of increased separation between BH in large disks described above and produces 614 mergers with the same disk lifetime.

\subsection{Initial Black Hole Spin Magnitude Distribution}
Here we explore the influence of the allowed initial BH spin magnitude distribution. We vary the width of the initial Gaussian distribution between $\sigma_\chi=[0.02,0.2]$ to reflect a range of natal spin magnitudes by setting \texttt{nsc\_bh\_spin\_dist\_sigma}.

\begin{figure*}
    \centering
    \includegraphics{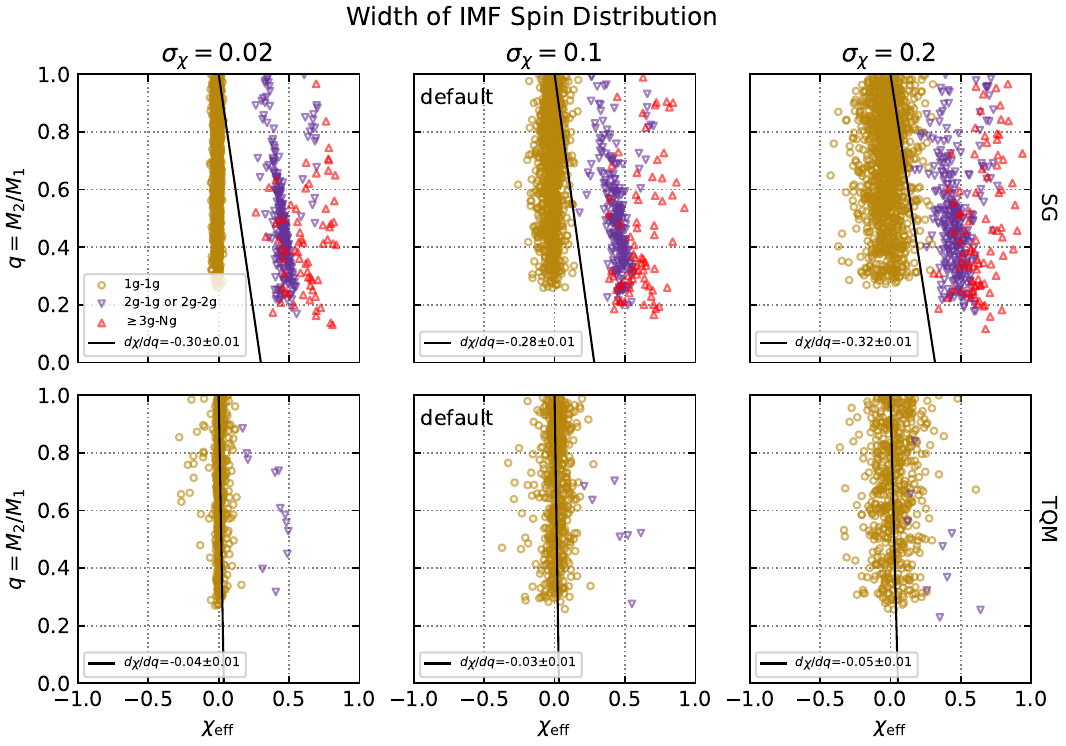}
    \caption{\textbf{Varying the width of the Gaussian distribution for initial BH spin magnitudes.}
    \textit{Top left:} SG disk with spin magnitude distribution $\sigma_\chi = 0.02$ (\texttt{sg\_0p02}).
    \textit{Top-center:} Default SG disk with $\sigma_\chi=0.1$ (\texttt{sg\_default}).
    \textit{Top-right:} SG disk with $\sigma_\chi = 0.2$ (\texttt{sg\_0p2}).
    \textit{Bottom row:} Same as top row but for a TQM disk model (\texttt{tqm\_0p02}, \texttt{tqm\_default}, \texttt{tqm\_0p2}).
    Narrower (wider) spin magnitude distributions primarily affect the 1g population by narrowing (widening) the distribution in $\Xeff$.
    Markers are as in \Fig{fig:default-imf}.
    See Table~\ref{tab:run-parameters} for unique setup parameters and Table~\ref{tab:popstats} for population statistics.}
    \label{fig:spin}
\end{figure*}

In \Fig{fig:spin} we see each test produces the usual population islands as the default setups. For SG disks, no significant change occurs in $\barq$ for any of the generational subpopulations (top left: \texttt{sg\_s0p02} and top right: \texttt{sg\_s0p20}). However, we see the spread in $\Xeff$ for the 1g population is greatly impacted and grows as the spin width increases from $\sigma_\chi= 0.02$ to $0.20$. The standard deviation on $\barXeff$(1g) reflects this trend, tightening to $-0.01\pm0.02$ when $\sigma_\chi = 0.02$ and broadening to $-0.02\pm0.10$ when $\sigma_\chi = 0.20$. While no statistically significant change is seen in $\barXeff$ for the hierarchical populations, tighter initial distributions of spin magnitude (smaller $\sigma_\chi$) cause their islands to be more concentrated, whereas a larger $\sigma_\chi$ increases their scatter.

The TQM model shows similar behavior as the SG disk when varying $\sigma_\chi$ (bottom left: \texttt{tqm\_s0p02} and bottom right: \texttt{tqm\_s0p20}). The $\barq$ of 1g mergers stays the same, but the variance on $\barXeff$ is correlated with increasing $\sigma_\chi$. For $\sigma_\chi=0.02,\ 0.10,\ 0.20$, $\barXeff$(1g)$=0.01\pm0.04,\ 0.01\pm0.07,\ 0.02\pm0.12$, respectively, but the number of hierarchical mergers is too low to make any robust claims regarding their means.

If the detected population has a component with a wide distribution of $\Xeff$ centered around $\Xeff=0$, for the AGN channel this implies 1g BH in NSCs have a broad range of initial spin values, possibly reflecting details of star formation and/or gas accretion in the galactic nucleus. If the measured effective spins follow narrow distributions (both centered on $\Xeff \sim 0$ and at higher $\Xeff$), then 1g BH in NSCs are likely born with a very narrow range of spins (close to zero). Constraints from future BBH catalogs (O4 and O5) on the width of the $\Xeff$ distribution will allow us to test models of BH birth and evolution in galactic nuclei. 

Changing $\sigma_\chi$ does not appear to significantly impact the expectation values for either disk model. But the intrinsic distributions in spin are likely $\sigma_\chi \leq 0.2$ in SG models since the standard deviation on $\Xeff$ for the 1g population resulting from this setting extends to negative $\Xeff$ about as far as \citet[]{Callister+21} report is observed. TQM models produce similar results for 1g mergers, but the number of hierarchical events remains limited.

\subsection{Impact of Retrograde Binaries}
\label{sec:retro-results}
\begin{figure*}
    \centering
    \includegraphics{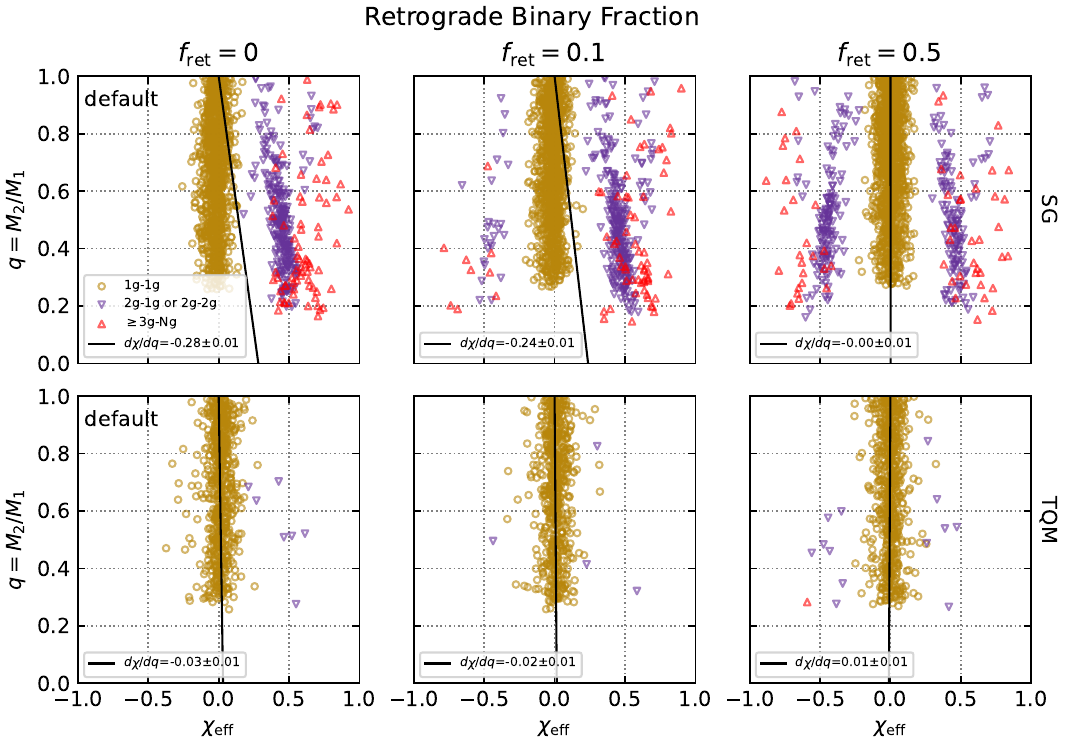}
    \caption{\textbf{Varying the retrograde BBH fraction.}
    \textit{Left column:} Default SG and TQM disk setups where no retrograde BBH form ($\fret=0$, \texttt{sg\_default}).
    \textit{Center column:}  Retrograde fraction is $\fret=0.1$ (\texttt{sg\_fr0p1}).
    \textit{Right column:} Half of BBH are retrograde ($\fret=0.5$)  (\texttt{sg\_fr0p5}).
    Increasing allowed retrograde fraction splits the hierarchical population to negative $\Xeff$ and flattens the slopes of the anti-correlation for SG-like disks. The TQM model (bottom row, \texttt{tqm\_default}, \texttt{tqm\_fr0p1}, \texttt{tqm\_fr0p5}) reflects this behavior but the hierarchical population still suffers from low numbers.
    Markers are as in \Fig{fig:default-imf}.
    See Table~\ref{tab:run-parameters} for unique setup parameters and Table~\ref{tab:popstats} for population statistics.}
    \label{fig:retro}
\end{figure*}

We now relax the default requirement for prograde BBH only and test a retrograde fraction $\fret=0.1,0.5.$ Retrograde BBH orbit retrograde around their center of mass, but prograde around the SMBH. Control this parameter in \texttt{McFACTS} with \texttt{fraction\_bin\_retro}.
A perfectly symmetric  distribution in $\Xeff$ around $\Xeff \sim 0$ consistent with is a $d\chi/dq$ slope of zero. This should by definition be the case for the $\fret=0.5$ (see \Fig{fig:retro}, where $d\chi/dq=0.0\pm0.01$).

The top row of \Fig{fig:retro} shows the results for increasing $\fret$ in a SG disk. In each case, the 1g population remains symmetric about $\Xeff=0$ without change from the fiducial setup. When $\vec{L}_{\rm bin}<0$ and $\chi_{1,2}>0$ for both BH, $\Xeff$ will be negative. This is also true when $q$ is sufficiently small and the more massive BH's spin is pointed opposite to $\vec{L}_{\rm bin}$. Increasing  $\fret=0$ (top left, $\sgdefault$) to $\fret=0.1$ (top center, \texttt{sg\_fr0p1}), primarily affects the mean effective spin of the hierarchical population, and we see a handful of these appear on the left side of the plot for $\fret=0.1$. For all hierarchical mergers, $\barXeff$ decreases with increasing retrograde fraction as more negative $\Xeff$ events occur, but the standard deviation increases drastically as a result. For $\fret=0$, $\barXeff$(2g)$=0.44\pm0.08$, but when $\fret=0.1$, $\barXeff$(2g)$=0.38\pm0.27$. The $\geq$3g population sees even greater uncertainty increases with $\barXeff(\geq$3g)$=0.61\pm0.13$ changing to $0.44\pm0.42$.

The change in $d\chi/dq$ reflects these shifts by drastically shifting towards negative $\Xeff$, flattening from $-0.28\pm0.01$ to $-0.24\pm0.01$. Even a few negative $\Xeff$ hierarchical events skew the measurement.

Increasing $\fret \rightarrow 0.5$ continues this trend (top right, \texttt{sg\_fr0p5}). The symmetry about $\Xeff=0$ we have seen in the 1g population now extends to the hierarchical mergers. Effectively $\fret=0.5$ is equivalent to a gas-free hierarchical merger environment. Since half of the hierarchical mergers now have $\Xeff<0$, the mean expectation value shifts to $\barXeff$(2g)$=-0.02\pm0.46$. There are no underlying differences to the hierarchical mergers compared to $\sgdefault$ (top left), which can be seen by taking the absolute value of the effective spin before calculating the mean $\overline{|\Xeff|}$(2g)$=0.45\pm0.08$ -- consistent with $\barXeff=0.44\pm0.08$ seen in \sgdefault. Similar behavior is reflected in the $\geq$3g population with $\barXeff(\geq$3g)$=0.06\pm0.63$ vs. $\overline{|\Xeff|}(\geq$3g)$=0.61\pm0.14$ when $\fret=0.5$ (see Table.~\ref{tab:popstats}).

The symmetry in the hierarchical population flattens the slope to $d\chi/dq=-0.24\pm0.01$ when $\fret=0.1$ and $0.00\pm0.01$ when $\fret=0.5$. Retrograde BBH are not expected to form this frequently in an AGN disk, but as one allows for more to occur by increasing $\fret$, the number of negative $\Xeff$ events increases, and the measured slope flattens.

The bottom row of \Fig{fig:retro} shows the results for the TQM model.
As in the SG model, changing $\fret$ to 0.1 (bottom center, \texttt{tqm\_fr0p1}) produces some $\Xeff<0$ hierarchical mergers, and $\fret=0.5$ (bottom right, \texttt{tqm\_fr0p5}) increases this number further, while $\barXeff({\rm 2g})$ decreases to $0.17\pm0.37$ for $\fret=0.1$ and to $-0.07\pm0.40$ when $\fret=0.5$. These shifts lead to decreasing slopes: $d\chi/dq=[-0.28\pm0.01, -0.24\pm0.01, 0.00\pm0.01]$. Though this disk model is plagued by low number statistics for hierarchical mergers, a small retrograde fraction still greatly affects the measurement of $d\chi/dq$.

\subsection{Black Hole Orbital Eccentricity}

In this section we explore the effects of changing the maximum initial orbital eccentricity $\emax$ of the embedded BH. Migration does not occur unless a BH has been circularized, defined when $e_{\rm crit}\leq0.01$. For eccentric prograde BH, mass accretion is retrograde, spinning down embedded BH. Thus, more eccentric BH that spend a longer time circularizing experience more spin-down before they can migrate and form BBH. Control this parameter in \texttt{McFACTS} with \texttt{disk\_bh\_orb\_ecc\_max\_init}

\subsubsection{SG disk: $\emax= 0.1, 0.3, 0.5$}

\begin{figure*}
    \centering
    \includegraphics{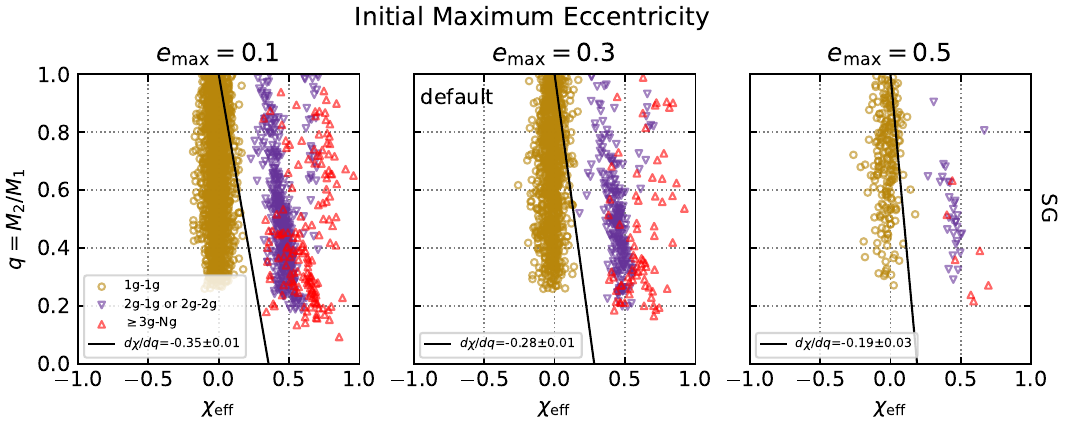}
    \caption{\textbf{Varying the initial maximum orbital eccentricity.}
    SG disk with NSC initial eccentricity capped at $e_{\rm max}=0.1$ (left, \texttt{sg\_e0p1}), 0.3 (center, \texttt{sg\_default}), and 0.5 (right, \texttt{sg\_0p5}).
    The number of mergers is inversely proportional to initial eccentricity.
    Markers are as in \Fig{fig:default-imf}.
    See Table~\ref{tab:run-parameters} for unique setup parameters and Table~\ref{tab:popstats} for population statistics.}
    \label{fig:ecc-sg}
\end{figure*}

\Fig{fig:ecc-sg} shows the results from a BH population with orbits constrained to be more circular ($\emax=0.1$, left panel, \texttt{sg\_e0p1}) or more eccentric ($\emax=0.5$, right panel, \texttt{sg\_e0p5}). Starting from lower $\emax$, the 1g population becomes saturated across the entire range of possible mass ratios, but no change occurs to the populations expectation values for $q$ or $\Xeff$. The same is true for the 2g and $\geq$3g populations. The slope does slightly steepen from $d\chi/dq=-0.28\pm0.01$ to $-0.35\pm0.01$ due to the relative increase in number of hierarchical mergers. Conversely, the number of mergers decreases in each group for a larger $\emax$, and the suppression of hierarchical mergers flattens the slope to $-0.19\pm0.03$.

Given a fixed $\tauAGN$, starting from orbits with a lower $\emax$ reduces time spent circularizing, jump starts the migration process, and increases the pool of BH available to form BBH at early times. Since more BH spend a longer time participating in mergers, the $\geq$3g hierarchical merger population is able to reach lower mass ratios and higher effective spins. 

\subsubsection{SG disk with thermal distribution: $\emax=0.7$}

\begin{figure*}
    \centering
    \includegraphics{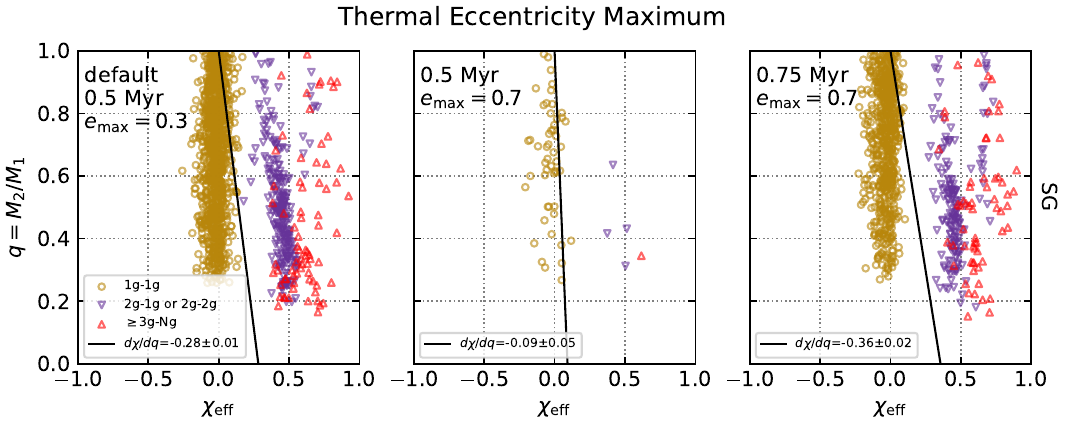}
    \caption{\textbf{Thermal Population with varying lifetimes.}
    \textit{Left:} Default SG disk with $\emax=0.3$ (\texttt{sg\_default}).
    \textit{Center:} SG disk interacting with a thermal NSC with $\emax=0.7$ (\texttt{sg\_e0p7}).
    \textit{Right:} Thermal NSC with SG disk lifetime increased to $\tauAGN=0.75$ Myr (\texttt{sg\_e0p7\_t0p75}).
    A more eccentric initial BH population takes longer to circularize which increases the time to produce mergers.
    Markers are as in \Fig{fig:default-imf}.
    See Table~\ref{tab:run-parameters} for unique setup parameters and Table~\ref{tab:popstats} for population statistics.}
    \label{fig:ecc-sg-thermal}
\end{figure*}

A thermal NSC is either one that has not had substantial recent AGN cooling activity or one that had sufficient time between active phases for dynamical encounters to erase the circularization caused by previous AGN activity and has median orbital eccentricity $\overline{e} \sim 0.7$. \Fig{fig:ecc-sg-thermal} shows an SG accretion disk interacting with a thermal BH population for $\tauAGN=0.5$ Myr (center panel, \texttt{sg\_e0p7}) suppresses almost all mergers. However, if $\tauAGN=0.7$ Myr (right panel, \texttt{sg\_e0p7\_t0p75}) there is sufficient time to damp embedded BH orbits to broadly match \sgdefault\ results. However, the 1g population qualitatively shifts towards negative $\Xeff$ --- though less so quantitatively (left panel: $\barXeff$(1g)$= -0.01\pm0.05$, right panel: $-0.02\pm0.06$) --- since accretion-driven spin down during a lengthy circularization period leads to initially highly eccentric BHs forming BBH with spins pointing opposite to the binary orbital angular momentum. This can be tested in LVK O4.

\subsubsection{TQM disk: $\emax = 0.015, 0.05, 0.1$}

\begin{figure*}
    \centering
    \includegraphics{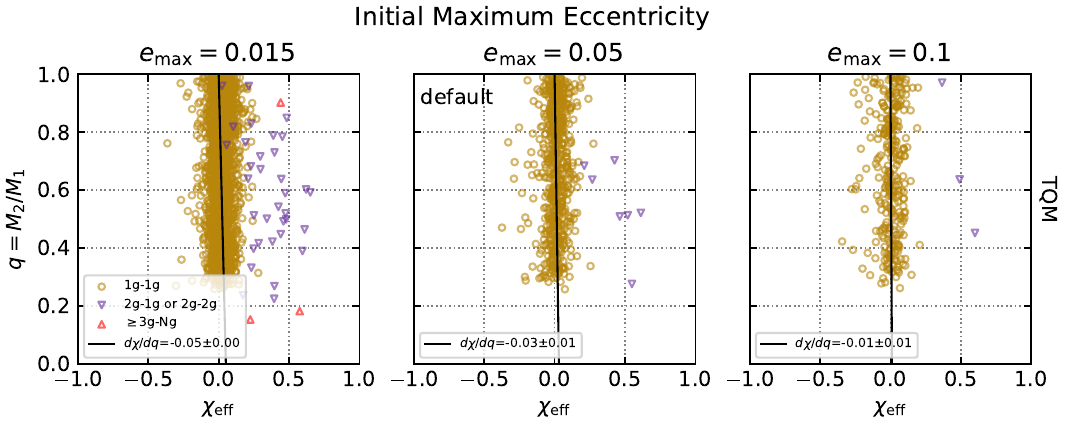}
    \caption{\textbf{Varying the initial maximum orbital eccentricity.}
    TQM disk with NSC initial eccentricity capped at $e_{\rm max}=0.015$ (left \texttt{tqm\_e0p015}), 0.05 (center \texttt{tqm\_default}), and 0.1 (right, \texttt{tqm\_e0p1}).
    The number of mergers is inversely proportional to initial eccentricity, and an extremely low eccentricity is required for the TQM model to produce an appreciable number of hierarchical mergers as opposed to the SG model.
    Markers are as in \Fig{fig:default-imf}.
    See Table~\ref{tab:run-parameters} for unique setup parameters and Table~\ref{tab:popstats} for population statistics.}
    \label{fig:ecc-tqm}
\end{figure*}

The default TQM setup uses a lower default eccentricity maximum $\emax = 0.05$ which we now vary between $\emax=0.015,0.1$. Examining \Fig{fig:ecc-tqm} shows that for $\emax=0.1$ (right panel, \texttt{tqm\_e0p1}) the hierarchical merger population is unchanged, and the 1g merger population drops by half. Even this small $\emax$ almost significantly suppresses mergers. Conversely, when $\emax = 0.015$ (left panel, \texttt{tqm\_e0p015})---only $\Delta e = 0.005$ above our eccentricity criterion for treating orbits as circular---we see for the first time in any TQM set-up, a distribution of hierarchical mergers that appears remotely similar to the SG model default. The 1g population is densely concentrated about $\Xeff=0$ with $\barXeff({\rm 1g})=0.02\pm0.06$ over the range of mass ratios $0.25 \lesssim q \leq 1$ with $\barq({\rm 1g})=0.68\pm0.21$. The hierarchical population is finally large enough for more robust expectation values of $\barXeff({\rm 2g})=0.35\pm0.16$ and $\barq({\rm 2g})=0.59\pm0.19$. However, there are over 60 times as many 1g mergers, and the anti-correlation's slope is flat at $d\chi/dq=-0.05\pm0.00$. 

\subsubsection{TQM disk with circular orbits: $\emax=0$}

The TQM model has been unable to produce any significant number of hierarchical mergers with a slightly eccentric initial NSC. Here we explore how $\tauAGN$ affects the $q$--$\Xeff$ plane, when the NSC starts from circularized orbits.

\begin{figure*}
    \centering
    \includegraphics{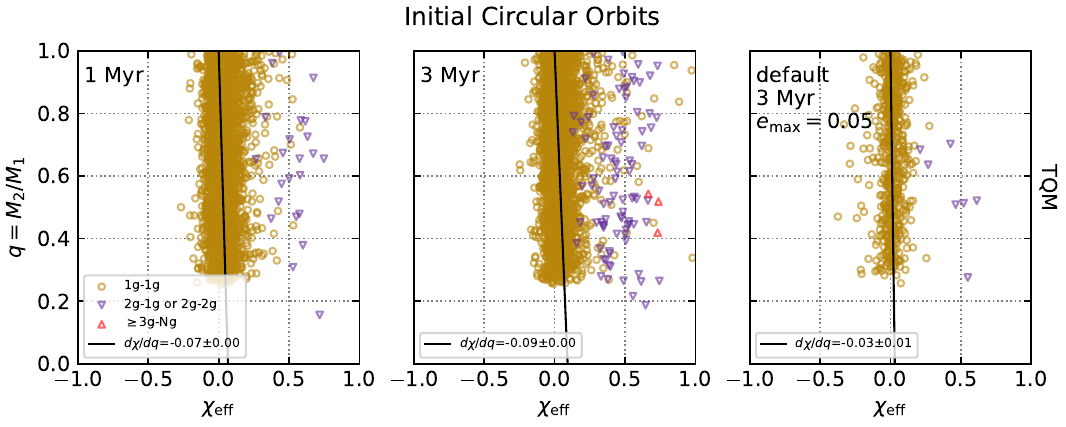}
    \caption{\textbf{Initially circular orbits with varying AGN lifetimes.}
    \textit{Left:} TQM disk evolved for $\tauAGN=1$ Myr (\texttt{tqm\_e0\_t1}).
    \textit{Center:} Setup same as \tqmdefault but with circular BH orbits ($\tauAGN=3$ Myr; \texttt{tqm\_e0}).
    \textit{Right:} Default TQM setup where $\tauAGN=3$ Myr and initial eccentricity is capped at $\emax=0.05$ (\texttt{tqm\_default}).
    Markers are as in \Fig{fig:default-imf}.
    See Table~\ref{tab:run-parameters} for unique setup parameters and Table~\ref{tab:popstats} for population statistics.}
    \label{fig:ecc-tqm-circular}
\end{figure*}

\Fig{fig:ecc-tqm-circular} shows the results from evolving the circular NSC ($\emax=0$) in a TQM disk for $\tauAGN=[1,3]$ Myr compared with our default setup. The only difference between the 3 Myr run (center panel, \texttt{tqm\_e0}) and the default setup (right panel) is the initial maximum eccentricity, yet already within 1 Myr the $\geq$2g population begins to populate more similar to \texttt{tqm\_e0p015}. At 1 Myr, the 1g population is still highly concentrated in effective spin, expected at $\barXeff({\rm 1g})=0.03\pm0.08$ and $\barq({\rm 1g})=0.68\pm0.21$. The mean effective spin for 2g hierarchical events is $\barXeff({\rm 2g})=0.53\pm0.12$, and the mean mass ratio is $\barq({\rm 2g})=0.63\pm0.20$. The hierarchical population does not affect anti-correlation's slope because there are almost 4000 1g mergers compared to the 23 2g mergers, causing the slope to flatten $d\chi/dq=-0.07\pm0.00$. After 3 Myr the number of 2g mergers increases to 89, and 3 $\geq$3g mergers occurred---the most in any TQM model. $\barXeff$(2g) moves to $0.45\pm0.14$ and is slightly larger at $0.71\pm0.03$ for $\geq$3g. The additional 69 $\geq$2g mergers and only $\sim$150 additional 1g mergers slightly shifts the slope to $d\chi/dq=-0.09\pm0.00$

By skipping the damping process for eccentric orbits, a circularized population of BH interacting within a TQM-like disk can immediately migrate, form binaries, and merge. By jump-starting the merger process, then number of hierarchical mergers produced within 1 Myr is four times larger than when the BH have even a tiny initial eccentricity ($\emax=0.015$). Though 3 Myr increases the hierarchical population, they primarily occur near $R=2\times10^4\,\Rg$, and far fewer higher generation ($\geq 3$g) mergers occur compared to SG disks which benefit from the trap. This behavior suggests TQM-like disks quickly exhaust the embedded 1g population, and are unable to continue merging higher generation BBH, because of slow (inefficient) migration in the outer disk. Thus, low density disks with dynamically cooled populations could contribute to the 1g population without significantly contributing to high mass BBH mergers.

\section{Discussion}
\label{sec:discussion}
One of the most intriguing results to emerge from the LVK O3 GW results was an apparent anti-correlation between BBH mass ratio ($q=M_{2}/M_{1}$) and the BBH effective spin ($\Xeff$) \citep{Callister+21}. At first glance this seems unexpected from most channels. The AGN channel was recognized as a possible source of such an anti-correlation simply because: a) the primary BH ($M_{1}$) in BBH should spend longer in the AGN disk and so, b) primary BH ($M_{1}$) spins should be biased towards alignment with the orbital angular momentum of the disk (and the BBH) \citep[e.g.][]{Mckernan+22:q-Xeff,Santini23, AlexD24}. Here we have demonstrated the form of the $q$--$\Xeff$ parameter space distribution as a function of different disk and model choices for the AGN channel.

First, and most importantly, our default choice of a SG disk run for 0.5 Myr with a $\gamma=2$ IMF seems to be a reasonable fit to the observed $q$--$\Xeff$ distribution (see \Fig{fig:default-imf} top left panel). \citet{Callister+21} find an overall anti-correlation fit of $d\chi/dq \sim -0.46^{+0.29}_{-0.28}$ for their distribution when they exclude GW190814. However, when they \emph{include} GW190814, they find $d\chi/dq=-0.28^{+0.31}_{-0.28}$ which matches $d\chi/dq=-0.28\pm0.01$ from our default SG model. This highlights an important point: GW190814 is difficult to reproduce commonly in any other channel, however in AGN (depending on choice of IMF) such low $q \sim 0.1$ events can occur at the few percent level \citep{Secunda19, McK20b,Hiromichi20}.

Second, it is clear that the AGN channel produces islands of population in $q$--$\Xeff$ parameter space. Evidently the 1g mergers tend (at least in dense SG-like disks) to be concentrated around $\Xeff \sim 0$. Higher generation hierarchical mergers ($\geq$2g) tend to live in islands at higher $\Xeff$. Indeed, the structure, location, and density of these population islands is what drives the $d\chi/dq$ in the AGN channel. For example, a steep IMF ($\gamma=2$) tends to drive $q \sim 0.5$ mergers, whereas a flatter IMF ($\gamma=1$) drives lower q mergers and drives the average ($d\chi/dq$) slope to more negative values (compare Fig.~\ref{fig:default-imf} top panels). Future LVK results (e.g. from O4) can constrain the presence (or not) of such islands. 

Third, while higher-density SG disks can reproduce the $d\chi/dq$ correlation, we find that low-density AGN disk models can only approach this correlation if the initial BH population has been dynamically cooled (compare e.g. Fig.~\ref{fig:default-imf} bottom panels and Fig.~\ref{fig:ecc-tqm-circular}). Indeed some level of dynamical cooling is also preferred in denser SG-like disks (see Fig.~\ref{fig:ecc-sg}). Only very small initial eccentricities for BH embedded in TQM-like disk models can reproduce the observed $d\chi/dq$ anti-correlation  \citep{Callister+21}. Lower surface density disk models like TQM take longer to circularize and migrate embedded BH, leading to lower merger rates than SG-like models. If we extend the disk lifetime ($\tauAGN$), we can drive a higher BBH merger rate, but at the expense of spin down due to retrograde accretion from the gas disk for an initially eccentric population (see Fig.~\ref{fig:time} bottom panels). So it is unlikely that low density disks are a significant contributor to the observed LVK rate among higher mass mergers unless their initial populations are dynamically cold (low $e$).

Fourth, as the width of BH initial spin distribution $\sigma_\chi$ increases, the width of the islands in $q$--$\Xeff$ parameter space increases (see gold, purple and red populations in  Fig.~\ref{fig:spin}). Larger values of $\sigma_\chi$ lead to a larger weighting of the 1g population to negative $\Xeff$. We constrain $\sigma_\chi \leq 0.2$ in the AGN channel from LVK observations.

Fifth, the gradient ($d\chi/dq$) in the AGN channel is also driven by the ratio ($\fret$) of retrograde/prograde BBH mergers. For $\fret=[0,0.1,5]$, the gradient $d\chi/dq=[-0.28\pm0.01,\ -0.24\pm0.01,\ 0.00\pm0.01]$ for our default model in an SG disk. As \citet{Callister+21} find $d\chi/dq = -0.28^{+0.31}_{-0.28}$ from O3, a low value of $\fret$ is preferred, but we cannot exclude $\fret \sim 0.5$. The value of $\fret$ for mergers in AGN disks remains an unsolved problem. The change in separation between BBH formation and merger is 6-7 orders of magnitude and the details of what happens during hardening in AGN remain unclear. BBH formation via GW capture in gas disks can be either prograde or retrograde \citep[e.g.][]{DeLaurentiis+23,Whitehead+24,Qian+24}. However, retrograde BBH may have their eccentricity pumped \citep{Li+22-bbh-evo,Calcino23}, and may therefore harden quickly towards merger. However, there is much debate as to whether retrograde BBH are preferentially ionized as their eccentricity is pumped \citep{Wang+22}, or whether they are flipped to prograde due to dynamical encounters \citep{Samsing22,McKF24}. The slope of the ($d\chi/dq$) anti-correlation that emerges from larger data sets (O4, O5) will allow us to directly test models of the retrograde fraction of BBH in the AGN channel.

Sixth, disk size impacts both the overall rate of BBH mergers and the $q$--$\Xeff$ distribution (see \Fig{fig:radius}). A small average disk size generates a plausible $q$--$\Xeff$ distribution but a far lower merger rate.

Seventh, more massive primaries tend to produce mergers with higher $\Xeff$ (see Fig.~\ref{fig:default-primarymass-xeff}) since they are more likely to be 2g or $\geq$3g and thus had their spins aligned with the disk. 

The observed LVK BBH rate of $\mathcal{R} \sim 25\,{\rm Gpc}^{-3} {\rm yr}^{-1}$ \citep{Abbot+2023:O3b} originates from a variety of formation channels, potentially including some fraction from AGN. From \citet{McKernan+24-McFACTS1} the merger rates for our default SG disk is $\mathcal{R} \sim 20\,{\rm Gpc}^{-3} \rm{yr}^{-1}$, whereas the default TQM rate is only $\mathcal{R} \sim 0.1\,{\rm Gpc}^{-3} \rm{yr}^{-1}$.  
SG-like AGN are good at making IMBH, driving low $q$ and high $\Xeff$ mergers, whereas low density/opacity, TQM-like disks are far less efficient producers of hierarchical merger events. As demonstrated, features in the $q$--$\Xeff$ plane due to AGN channel BBH mergers depend strongly on the disk model, so LVK O4 results will allow us to constrain possible AGN disk models.

\section{Conclusion}
\label{sec:conclusion}
We use the new public, open-source, reproducible code \texttt{McFACTS} to test the $q$--$\Xeff$ distribution for the LVK AGN channel. Broadly we find islands of structure in $q$--$\Xeff$ parameter space as a function of BH generation, spin and eccentricity distributions and disk models and lifetime. The observed $q$--$\Xeff$ anti-correlation can be well reproduced by a dense and relatively short-lived model AGN disk with some dynamical cooling. 
Tighter constraints on the $q$--$\Xeff$ distribution from future LVK BBH catalogs (O4 and O5) will allow us to strongly constrain key AGN disk and NSC properties.

\begin{acknowledgments}
We thank the referee for a thorough report that helped improve the code and paper.
HEC, BM \& KESF are supported by NSF AST-2206096. BM \& KESF are also supported by Simons Foundation Grant 533845 as well as Simons Foundation sabbatical support. The Flatiron Institute is supported by the Simons Foundation.
ROS  gratefully acknowledges support from NSF awards NSF PHY-1912632, PHY-2012057, PHY-2309172, AST-2206321, and the Simons Foundation.
VD is supported by an appointment to the NASA Postdoctoral Program at the NASA Goddard Space Flight Center administered by Oak Ridge Associated Universities under contract NPP-GSFC-NOV21-0031.
KN thanks the LSST-DA Data Science Fellowship Program, which is funded by LSST-DA, the Brinson Foundation, and the Moore Foundation; her participation in the program has benefited this work. 
\end{acknowledgments}

\software{Astropy, \citep{2013A&A...558A..33A,2018AJ....156..123A,2022ApJ...935..167A}, pAGN \citep{Gangardt+24-pAGN}, NumPy \citep{harris2020array}, SciPy \citep{2020SciPy-NMeth}, Matplotlib \citep{Hunter_2007}.
This work uses McFACTS Version 0.3.0, which is available on Zenodo at \href{https://doi.org/10.5281/zenodo.14422240}{10.5281/zenodo.14422240}.}

\bibliography{refs}{}
\bibliographystyle{aasjournal}

\movetabledown=3.5cm
\begin{rotatetable*}
\begin{deluxetable*}{r|c|c|ccc|ccc|ccc}
\label{tab:popstats}
\tabletypesize{\scriptsize}
\tablewidth{0pt} 
\tablecaption{Statistical estimates of the merger events.}
\tablehead{
\colhead{Run} & \colhead{Fig.} & \colhead{$d\chi/dq$} & \multicolumn{3}{c}{1g-1g Mergers} & \multicolumn{3}{c}{2g-[1,2]g Mergers} & \multicolumn{3}{c}{$\geq$3g-Ng Mergers} \\
\colhead{} & \colhead{} & \colhead{} & \colhead{N} & \colhead{$\barq$} & \colhead{$\barXeff$} & \colhead{N} & \colhead{$\barq$} & \colhead{$\barXeff$} & \colhead{N} & \colhead{$\barq$} & \colhead{$\barXeff$}
}

\startdata
\texttt{sg\_default} & \ref{fig:default-imf}, \ref{fig:default-radius-time}, \ref{fig:default-primarymass-xeff} & $-0.28\pm0.01$ & 1093 & $ 0.67\pm0.19$ & $-0.02\pm0.05$ & 244 & $ 0.51\pm0.18$ & $ 0.44\pm0.08$ & 73 & $ 0.45\pm0.22$ & $ 0.61\pm0.13$ \\
\texttt{tqm\_default} & \ref{fig:default-imf}, \ref{fig:default-radius-time}, \ref{fig:default-primarymass-xeff} & $-0.03\pm0.01$ & 606 & $ 0.68\pm0.20$ & $ 0.01\pm0.07$ & 7 & $ 0.55\pm0.14$ & $ 0.44\pm0.14$ & 0 & --- & --- \\
\hline
\texttt{sg\_g1} & \ref{fig:default-imf} & $-0.33\pm0.01$ & 1925 & $ 0.63\pm0.21$ & $-0.01\pm0.06$ & 556 & $ 0.50\pm0.21$ & $ 0.46\pm0.10$ & 207 & $ 0.44\pm0.22$ & $ 0.63\pm0.16$ \\
\texttt{tqm\_g1} & \ref{fig:default-imf} & $-0.05\pm0.01$ & 683 & $ 0.65\pm0.23$ & $ 0.01\pm0.07$ & 18 & $ 0.55\pm0.27$ & $ 0.37\pm0.12$ & 0 & --- & --- \\
\hline
\texttt{sg\_t0p25} & \ref{fig:time}, \ref{fig:time-mass} & $-0.05\pm0.07$ & 20 & $ 0.70\pm0.19$ & $-0.00\pm0.06$ & 1 & $ 0.63\pm0.00$ & $ 0.47\pm0.00$ & 0 & --- & --- \\
\texttt{sg\_t0p75} & \ref{fig:time}, \ref{fig:time-mass} & $-0.59\pm0.01$ & 2246 & $ 0.69\pm0.19$ & $-0.02\pm0.05$ & 1004 & $ 0.51\pm0.18$ & $ 0.47\pm0.09$ & 756 & $ 0.44\pm0.22$ & $ 0.67\pm0.15$ \\
\texttt{tqm\_t2p5} & \ref{fig:time}, \ref{fig:time-mass} & $-0.05\pm0.01$ & 572 & $ 0.67\pm0.21$ & $ 0.02\pm0.06$ & 6 & $ 0.48\pm0.24$ & $ 0.36\pm0.20$ & 0 & --- & --- \\
\texttt{tqm\_t5} & \ref{fig:time}, \ref{fig:time-mass} & $-0.04\pm0.01$ & 620 & $ 0.67\pm0.21$ & $ 0.00\pm0.10$ & 17 & $ 0.52\pm0.23$ & $ 0.36\pm0.21$ & 0 & --- & --- \\
\hline
\texttt{sg\_r2e4} & \ref{fig:radius}, \ref{fig:radius-mass} & $-0.45\pm0.02$ & 465 & $ 0.60\pm0.19$ & $ 0.02\pm0.06$ & 187 & $ 0.46\pm0.17$ & $ 0.48\pm0.08$ & 96 & $ 0.47\pm0.25$ & $ 0.62\pm0.15$ \\
\texttt{sg\_r7e4} & \ref{fig:radius}, \ref{fig:radius-mass} & $-0.25\pm0.03$ & 198 & $ 0.70\pm0.19$ & $-0.02\pm0.05$ & 30 & $ 0.54\pm0.20$ & $ 0.45\pm0.10$ & 11 & $ 0.42\pm0.23$ & $ 0.63\pm0.10$ \\
\texttt{tqm\_r2e4} & \ref{fig:radius}, \ref{fig:radius-mass} & $-0.13\pm0.02$ & 185 & $ 0.70\pm0.21$ & $ 0.04\pm0.07$ & 7 & $ 0.71\pm0.17$ & $ 0.43\pm0.09$ & 0 & --- & --- \\
\texttt{tqm\_r7e4} & \ref{fig:radius}, \ref{fig:radius-mass} & $-0.05\pm0.01$ & 600 & $ 0.66\pm0.21$ & $ 0.01\pm0.07$ & 12 & $ 0.65\pm0.20$ & $ 0.36\pm0.09$ & 0 & --- & --- \\
\hline
\texttt{sg\_s0p02} & \ref{fig:spin} & $-0.30\pm0.01$ & 1234 & $ 0.68\pm0.20$ & $-0.01\pm0.02$ & 262 & $ 0.50\pm0.19$ & $ 0.46\pm0.07$ & 67 & $ 0.42\pm0.18$ & $ 0.62\pm0.14$ \\
\texttt{sg\_s0p20} & \ref{fig:spin} & $-0.32\pm0.01$ & 1263 & $ 0.69\pm0.19$ & $-0.02\pm0.10$ & 299 & $ 0.50\pm0.19$ & $ 0.47\pm0.10$ & 87 & $ 0.47\pm0.22$ & $ 0.64\pm0.13$ \\
\texttt{tqm\_s0p02} & \ref{fig:spin} & $-0.04\pm0.01$ & 595 & $ 0.70\pm0.20$ & $ 0.01\pm0.04$ & 12 & $ 0.61\pm0.17$ & $ 0.37\pm0.12$ & 0 & --- & --- \\
\texttt{tqm\_s0p20} & \ref{fig:spin} & $-0.05\pm0.01$ & 586 & $ 0.67\pm0.21$ & $ 0.02\pm0.12$ & 9 & $ 0.47\pm0.19$ & $ 0.32\pm0.16$ & 0 & --- & --- \\
\hline
\texttt{sg\_fr0p1} & \ref{fig:retro} & $-0.24\pm0.01$ & 1144 & $ 0.68\pm0.19$ & $-0.01\pm0.05$ & 265 & $ 0.50\pm0.19$ & $ 0.38\pm0.27$ & 69 & $ 0.43\pm0.22$ & $ 0.44\pm0.42$ \\
\texttt{sg\_fr0p5} & \ref{fig:retro} & $0.00\pm0.01$ & 1177 & $ 0.70\pm0.19$ & $ 0.00\pm0.06$ & 255 & $ 0.51\pm0.19$ & $-0.02\pm0.46$ & 61 & $ 0.46\pm0.20$ & $ 0.06\pm0.63$ \\
\texttt{tqm\_fr0p1} & \ref{fig:retro} & $-0.02\pm0.01$ & 580 & $ 0.67\pm0.21$ & $ 0.01\pm0.07$ & 4 & $ 0.51\pm0.19$ & $ 0.17\pm0.37$ & 0 & --- & --- \\
\texttt{tqm\_fr0p5} & \ref{fig:retro} & $ 0.01\pm0.01$ & 601 & $ 0.67\pm0.21$ & $0.00\pm0.07$ & 13 & $ 0.50\pm0.15$ & $-0.07\pm0.40$ & 1 & $ 0.28\pm0.00$ & $-0.59\pm0.00$ \\
\hline
\texttt{sg\_e0p1} & \ref{fig:ecc-sg} & $-0.35\pm0.01$ & 1778 & $ 0.68\pm0.20$ & $-0.01\pm0.05$ & 464 & $ 0.50\pm0.19$ & $ 0.46\pm0.08$ & 156 & $ 0.45\pm0.22$ & $ 0.62\pm0.14$ \\
\texttt{sg\_e0p5} & \ref{fig:ecc-sg} & $-0.19\pm0.03$ & 231 & $ 0.71\pm0.17$ & $-0.02\pm0.06$ & 30 & $ 0.53\pm0.14$ & $ 0.44\pm0.07$ & 7 & $ 0.37\pm0.14$ & $ 0.54\pm0.10$ \\
\texttt{sg\_e0p7} & \ref{fig:ecc-sg-thermal} & $-0.09\pm0.05$ & 51 & $ 0.65\pm0.18$ & $-0.03\pm0.07$ & 4 & $ 0.45\pm0.12$ & $ 0.45\pm0.06$ & 1 & $ 0.35\pm0.00$ & $ 0.62\pm0.00$ \\
\texttt{sg\_e0p7\_t0p75} & \ref{fig:ecc-sg-thermal} & $-0.36\pm0.02$ & 631 & $ 0.71\pm0.18$ & $-0.03\pm0.06$ & 161 & $ 0.52\pm0.18$ & $ 0.46\pm0.09$ & 61 & $ 0.50\pm0.20$ & $ 0.63\pm0.13$ \\
\hline
\texttt{tqm\_e0p015} & \ref{fig:ecc-tqm} & $-0.05\pm0.00$ & 2041 & $ 0.68\pm0.21$ & $ 0.02\pm0.06$ & 34 & $ 0.59\pm0.19$ & $ 0.35\pm0.16$ & 3 & $ 0.41\pm0.35$ & $ 0.41\pm0.14$ \\
\texttt{tqm\_e0p1} & \ref{fig:ecc-tqm} & $-0.01\pm0.01$ & 300 & $ 0.66\pm0.21$ & $ 0.00\pm0.08$ & 3 & $ 0.69\pm0.21$ & $ 0.49\pm0.10$ & 0 & --- & --- \\
\texttt{tqm\_e0\_t1} & \ref{fig:ecc-tqm-circular} & $-0.07\pm0.00$ & 3901 & $ 0.68\pm0.21$ & $ 0.03\pm0.08$ & 23 & $ 0.63\pm0.20$ & $ 0.53\pm0.12$ & 0 & --- & --- \\
\texttt{tqm\_e0} & \ref{fig:ecc-tqm-circular} & $-0.09\pm0.00$ & 4066 & $ 0.68\pm0.21$ & $ 0.03\pm0.09$ & 89 & $ 0.59\pm0.21$ & $ 0.45\pm0.14$ & 3 & $ 0.49\pm0.05$ & $ 0.71\pm0.03$ \\
\enddata
\tablecomments{We report the $d\chi/dq$ slope of a line fit to all mergers anchored through the point $(\Xeff,q)=(0,1)$, the number of mergers, the mean mass ratio $\barq$, mean effective spin $\barXeff$, and relevant standard deviations, for 1g-1g, 2g-Ng (2g-1g and 2g-2g), and $\geq$3g-Ng merger events in the associated figure.}
\label{tab:runs}
\end{deluxetable*}
\end{rotatetable*}

\end{document}